\documentclass{article}

\usepackage{arxiv}

\usepackage[utf8]{inputenc} 
\usepackage[T1]{fontenc}    
\usepackage{hyperref}       
\usepackage{url}            
\usepackage{booktabs}       
\usepackage{amsfonts}       
\usepackage{nicefrac}       
\usepackage{microtype}      
\usepackage{lipsum}
\usepackage{graphicx}
\usepackage{booktabs}
\usepackage{multirow}
\usepackage{siunitx}

\title{Taxonomy of Centralization in Public Blockchain Systems: A Systematic Literature Review}

\author{
  Ashish Rajendra Sai \\
  Lero-Irish Software Research Centre\\
  University of Limerick\\
  Limerick, Ireland \\
  \texttt{17053145@studentmail.ul.ie} \\
   \And
 Jim Buckley \\
  Lero-Irish Software Research Centre\\
  University of Limerick\\
  Limerick, Ireland \\
  \And 
  Brian Fitzgerald\\
  Lero-Irish Software Research Centre\\
  University of Limerick\\
  Limerick, Ireland \\
  \And 
  Andrew Le Gear\\
  Horizon Globex Ireland DAC\\
  Nexus Center, University of Limerick\\
  Limerick, Ireland \\
}

\begin{document}
\maketitle

\begin{abstract}
Bitcoin introduced delegation of control over a monetary system from a select few to all who participate in that system. This delegation is known as the decentralization of controlling power and is a powerful security mechanism for the ecosystem. After the introduction of Bitcoin, the field of cryptocurrency has seen widespread attention from industry and academia, so much so that the original novel contribution of Bitcoin, i.e., decentralization, may be overlooked, due to decentralizations' assumed fundamental existence for the functioning of such crypto-assets. 
  However, recent studies have observed a trend of increased centralization in cryptocurrencies such as Bitcoin and Ethereum. As this increased centralization has an impact the security of the blockchain, it is crucial that it is measured, towards adequate control. This research derives an initial taxonomy of centralization present in decentralized blockchains through rigorous synthesis using a systematic literature review. This is followed by iterative refinement through expert interviews. We systematically analyzed 89 research papers published between 2009 and 2019. Our study contributes to the existing body of knowledge by highlighting the multiple definitions and measurements of centralization in the literature. We identify different aspects of centralization and propose an encompassing taxonomy of centralization concerns. This taxonomy is based on empirically observable and measurable characteristics. It consists of 13 aspects of centralization, classified over six architectural layers: Governance, Network, Consensus, Incentive, Operational, and Application. We also discuss how the implications of centralization can vary depending on the aspects studied.
   We believe that this review and taxonomy provides a comprehensive overview of centralization in decentralized blockchains involving various conceptualizations and measures.  
\end{abstract}


\section{Introduction \label{1introduction}}
 
Since the introduction of Bitcoin in 2009, blockchain technology has seen a proliferation of scholarly articles investigating the potential and limitations of the technology \cite{beck2017blockchain,yli2016current,mattila2016blockchain,androulaki2018hyperledger,wust2018you,zheng2017overview,beck2018governance,walport2016distributed,davidson2016economics,he2017survey}. Control over the system is a focal point in a significant proportion of these studies, as this either enhances or restricts the usability of blockchain \cite{cong2019decentralized,gervais2014bitcoin,SaiBuckleyLeGear2019,beck2017blockchain,gencer2018decentralization,beikverdi2015trend,azouvi2018egalitarian,kwon2019impossibility,zhang2019security,mattila2016blockchain,zheng2017overview,judmayer2017blocks,baliga2017understanding,104,43}. Indeed, removing central control from the monetary system while continuing to ensure security has been considered as a core novel contribution of Bitcoin \cite{bonneau2015sok}. There are three main types of blockchain solutions based on the type of control mechanism used: public, private, and consortium \cite{zheng2017overview}.

In public blockchains, such as Bitcoin and Ethereum, every participant in the network contributes to the control mechanism, agreeing on a single state of the data without the need for a trusted third party. All  participants can read and write to this single state without any authorization \cite{guegan:halshs-01524440}. This consensus is achieved under the assumption of delegation of power of control, and the assumption that the majority of the network participants remain honest i.e., non-malicious. This delegation of  power of control is often referred to as decentralization. 

Contrary to the decentralized nature of a public blockchain, private and consortium blockchains, such as Hyperledger, tend to impose constraints on participants by including trusted entities in the system \cite{androulaki2018hyperledger}. These constraints can also include limitations on read and write permissions of participants \cite{guegan:halshs-01524440}. Based on the sensitivity of the information processed by the blockchain, practitioners may decide to adopt one of these controlling mechanisms \cite{meijer2018governance,wust2018you,peck2017blockchain}. This decision is potentially problematic, e.g., in the case of a practitioner who decides to use a public blockchain for decentralizing the control. As reported by Sai et al. (2019) \cite{SaiBuckleyLeGear2019}, decentralization in public blockchain is not a fundamental given by design, but a non-deterministic and probabilistic guarantee provided by clever integration of cryptography, distributed systems, and incentive engineering. 

The removal of trusted entities from a distributed system makes a public blockchain attractive to numerous potential users in academia and industry \cite{mattila2016blockchain}. Public blockchain-based cryptocurrencies have a market capitalization of over \$200 million \cite{sai2019privacy}, making the platform a lucrative target for malicious actors. The majority of these blockchains use decentralization as a security mechanism. In a decentralized system, the malicious actor would need to compromise half of the consensus power before causing significant harm to the system \cite{karame2012double}. Because of this interplay between decentralization and security, it is highly desirable to have a high degree of decentralization in public blockchains.  The security of a public blockchain has been thoroughly investigated in research \cite{bonneau2015sok,halpin2017introduction,karame2016bitcoin,karame2016security}. For example, Bitcoin has been reported as secure, subject to its adherence to the honest majority assumption, with notable exceptions such as selfish mining attacks \cite{sapirshtein2016optimal} where the attacker only needs to control over 26 \% of the network.

Even though the initial implementation of Bitcoin was able to circumvent the need for centralization in the system, new avenues of centralization are surfacing \cite{gervais2014bitcoin}. Numerous studies have reported various forms of centralization in Bitcoin and other decentralized cryptocurrency systems \cite{azouvi2018egalitarian,beikverdi2015trend,gencer2018decentralization,gervais2014bitcoin}. These reports of a trend towards centralization have raised security concerns as the security guarantee of a public blockchain is inherently dependent on the honest majority assumption \cite{SaiBuckleyLeGear2019}. Trusting the probabilistic security guarantees of a public blockchain has often been identified as a barrier to entry in the ecosystem \cite{iansiti2017truth}. Understanding more fully the security implication of centralization will aid the process of public blockchain adoption.  

The threats of centralization range well beyond security into adoption, and even crypto-economics \cite{conti2018survey}. The decentralized nature of bitcoin permits the uncensored execution of transactions in the payment system irrespective of political or geographical associations. Centralization may threaten the uncensored nature of the decentralized blockchain. Thus, it is crucial for the security and, consequently, the utility of public blockchain systems that they remain adequately decentralized.

Given the significance of decentralization, several studies have analyzed technical aspects \cite{beikverdi2015trend,gencer2018decentralization,gervais2014bitcoin} as well as social constructs of decentralization \cite{azouvi2018egalitarian}. By far, the most commonly measured aspects of centralization is the consensus power concentration \cite{azouvi2018egalitarian,beikverdi2015trend,gencer2018decentralization,gervais2014bitcoin,kwon2019impossibility}. In a Proof-of-Work based blockchain solution, the individual participants' consensus power is defined by their computational power in proportion to the total computational power of the network. However, this measurement mechanism is only useful in determining the present state of the computational power portions of the network. It fails to capture the multitude of factors that may constitute the overall centralization of the system, such as system governance \cite{beck2018governance}, wealth concentration \cite{chohan2019cryptocurrencies}, and geographic distribution of participants \cite{gencer2018decentralization}. 

 To better understand the semantics of decentralization in blockchain, we intend to measure it on all building blocks of the public blockchain. As reported by Wang et al. (2017) \cite{wang2017internal}, the governance structure of the blockchain can have a profound impact on the operations of a public blockchain but is often overlooked as a potential source of centralization. The issues caused by centralization of governance include the long-discussed issue of block size in Bitcoin \cite{caffyn2015bitcoin} and specific instances of unilateral decision making regarding forks in Ethereum \cite{wirdum2016rejecting}. Consequently, we reason that we need a vocabulary to discuss and measure centralization in a more holistic manner.

To allow for such modular measurement of centralization, we review the generic architecture \cite{zhang2019security} of blockchain and use it to identify potential avenues of centralization, via a literature review of the field. Focusing on the generic architecture enables us to capture centralization-causing factors that are not implementation-specific, i.e., the same model may be used for both Bitcoin and Ethereum.  We also use the generic architecture to partition different centralization concerns into architectural categories such as consensus, network, and application. This abstraction allows us to organize and observe centralization holistically. Thus, in this work, we present the first in-depth analysis of centralization in blockchains to assess the following questions:

\begin{center}
\textit{RQ1: What are the different aspects of centralization in public blockchains? } \\
\textit{RQ2: How can centralization be adequately measured in a decentralized blockchain instance?}
\end{center}

\begin{figure}[h]
  \centering
  \includegraphics[scale=0.40]{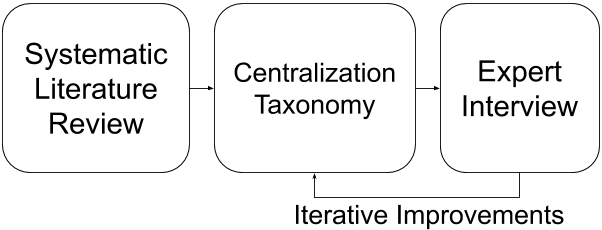}
  \caption{Methodology}
\end{figure}

To study decentralization in blockchain, we coded and analyzed the content of relevant blockchain literature. We chose ten years subsequent to the publication of the original Bitcoin white paper \cite{nakamoto2008bitcoin}. The survey process was primarily driven by the guidelines provided by Kitchenham et al. (2004) \cite{kitchenham2004procedures}. In adherence to the guidelines, we conducted a five-step systematic literature review consisting of \textit{Search, Selection, Quality Assessment, Data Extraction, and Analysis}. This systematic literature review produced the final article pool of 89 articles. These final articles, partitioned by architectural components, form the basis of the taxonomy proposed in this review.

Following the development of the taxonomy, we interviewed industrial and academic experts in the blockchain domain to establish the completeness of the taxonomy and to assess any redundant or less relevant components of the taxonomy. This consisted of ten expert interviews: four academic researchers and six industry experts. It resulted in an iterative refinement of the taxonomy. 

The paper makes the following contributions:

\begin{itemize}
  \item We systematically review the existing literature to document the different aspects of centralization in public blockchains (Section 3).
  \item We outline the different techniques employed in the literature to measure centralization (Section 4). 
  \item We manifest the findings of our review in a conceptual taxonomy that encompasses both categorization and measurement of different aspects of centralization in public blockchains (Section 4).
  \item We illustrate the relevance and utility of this taxonomy by presenting the centralization state of the two most prominent blockchain instances: Bitcoin and Ethereum, based on this taxonomy (Section 5). We also discuss how the adverse impact of centralization varies depending on aspects (Section 6).
  \item We identify research gaps specifically with regards to the lack of non-Bitcoin-specific centralization investigations. We also report on the lack of objective metrics for some centralization causing factors.
\end{itemize} 

\section{Background \label{2background}}
The term blockchain is often used as a generic descriptor for the broader field of Distributed Ledger Technologies \cite{great2016distributed}. Distributed ledger technology refers to the distributed computing networks that record, share, and synchronize data across many participants. More specifically, Blockchain is a type of data structure used to record data on these distributed computing networks. It is a chronologically linked list of data packets received by the participants within a predefined time period. These blocks are connected in a chronological order to form a chain of blocks. The link between these blocks is secured by the use of a computationally hard cryptographic hash function based puzzle \cite{nakamoto2008bitcoin}. As the chain of blocks grows, the difficulty involved in recalculating the puzzles also grows to make any alteration to past data expensive. This growth in difficulty leads to a deterministic guarantee of data immutability.  


The participants of the blockchain-based network have to reach consensus on a single state of this append-only structure.  Blockchain-based systems utilize a peer-to-peer distributed system with a clever incentive mechanism \cite{baliga2017understanding} to accomplish this consistency of data in an unconstrained distributed environment. Proof-of-work (PoW) and Proof-of-stake (PoS) are two prominent examples of consensus mechanisms used in blockchain-based systems. In PoW, the participants are expected to perform computationally expensive operations to solve a puzzle. The first participant to solve and propagate the solution to a majority of the network is rewarded. PoW is often criticized for the extensive use of electricity \cite{o2014bitcoin}. This issue of electricity usage is addressed in PoS, where the reward distribution is based on the monetary assets of the participants \cite{35}. Other notable consensus algorithms include Proof-of-Authority, Proof of Elapsed Time, and Delegated Proof-of-Stake; we refer the reader to \cite{mingxiao2017review} for an in-depth review of consensus algorithms.

As discussed earlier, based on the type of consensus mechanism deployed and the constraints imposed, we can segment blockchain-based systems in three broad categories: Public, Private, and Consortium. In private and consortium-based blockchain systems, the participation in consensus is limited to users approved by a trusted authority. However, in Public blockchain systems, the participation in consensus is open to any individual with appropriate computing and networking capabilities. This unconstrained access to controlling power for all participants in the network is referred to as decentralization. Bitcoin and other public blockchains establish consensus on the blockchain through a decentralized, pseudonymous protocol. This protocol can be considered a core innovation and possibly the most crucial ingredient to the success of public blockchains \cite{bonneau2015sok}.

\subsection{Decentralization and Public Blockchain\label{2.2background}}
 Decentralization is an essential property of public Blockchain systems where participants can read, write data, and contribute to consensus without authorization \cite{davidson2016economics}. In this subsection, we review the existing discussion around decentralization in the blockchain.

Consensus on the state of data in a public blockchain is attained by the acceptance of a valid block by the network in a predefined time interval. To deter malicious participants from accepting fraudulent blocks, the majority of the control must be decentralized. This decentralization of control ensures that the blockchain is secure from malicious participants as long as the majority of the network remains honest. This interplay of security and decentralization makes it fundamental that the system remains decentralized. 

A survey paper by He et al. (2017) \cite{he2017survey} identifies decentralization, among other features, as a prominent reason to adopt blockchain technology for business applications. This view is supported by numerous studies which demonstrate the application of decentralized Blockchains to the liberalization of financial asset management \cite{guo2016blockchain}, the Internet of Things \cite{panarello2018blockchain,zhu2019applications}, healthcare \cite{dwivedi2019decentralized} and smart cities \cite{xie2019survey}. The extent of literature surveyed by these review articles demonstrates the significance of decentralization in blockchain applications.

As decentralization is core to the secure functioning of public blockchains, it may be taken as a fundamental given. This assumed association between decentralization and public blockchains may be a vulnerability that malicious actors attack.  Security research on the blockchain has focused on the assumption of an honest majority. A survey paper by Li et al. (2017) \cite{li2017survey} identifies the centralization of consensus power as a significant security threat to that network. Centralization of consensus power is intrinsic to attacks on the public blockchain, such as the 51\% attack \cite{bradbury2013problem} and Selfish Mining \cite{sapirshtein2016optimal}. 

In the 51\% attack, the attacker is assumed to have gained control of more than half of the consensus power, which can then be used to enter fraudulent transactions in the blockchain. Unlike the 51\% attack, in selfish mining, the attacker only needs to control 26 \% consensus power to cause harm to the network \cite{SaiBuckleyLeGear2019}. More detail on the security of blockchains is provided in Zhang et al. (2019) \cite{zhang2019security}.

Studying blockchain as from solely a technical perspective may be misleading due to the inherent socio-technical nature of the blockchain \cite{de2019fragility}. As the study of centralization in public blockchain is still fragmented, current conceptual models, such as security and privacy models, do not provide adequate insights. To overcome this limitation, we devise a novel centralization taxonomy focusing on the different architectural layers of blockchain to categorize centralization concerns. We employ a two-step research approach, first conducting a systematic literature review to construct a taxonomy of centralization, and refine this further through expert interviews.


\subsection{Architecture of Public Blockchains\label{2.1background}}

The first public blockchain, Bitcoin, incorporated the blockchain data structure and consensus mechanism in-depth, but omitted any formalization of the networking structure \cite{nakamoto2008bitcoin}. Since the introduction of Bitcoin, numerous attempts have been made to describe the structure of public blockchains more formally.

Some of these attempts have been aspect-specific with a microscopic focus on one or a few components of the blockchain. For example, Garay et al. (2015) \cite{garay2015bitcoin} describe the architecture of blockchain in terms of consensus mechanisms and participants of the network. Another notable description of blockchain architecture is given by Gervais et al. (2016) \cite{gervais2016security}, who focus on security and scalability by describing consensus and a peer-to-peer network. 

Since the aim of our review is to analyze public blockchains more holistically to capture the factors causing centralization, we adhere to a more generic description of blockchain used by Zhu et al. (2019) \cite{zhu2019applications} and Zhang et al. (2019)\cite{zhang2019security}. In this generic description, the authors propose a layered architecture of blockchain. As a blockchain is a peer-to-peer distributed network, it is intuitive that blockchain systems will share many similarities with a generic, distributed computing architecture, such as the traditional OSI layered model of a network \cite{briscoe2000understanding}. 

\begin{figure}[h]
  \centering
  \includegraphics[scale=0.4]{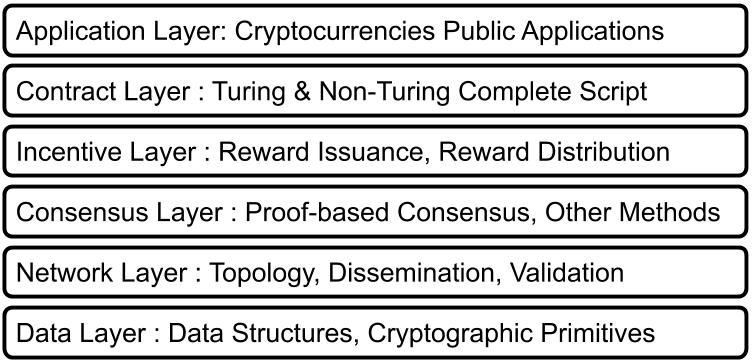}
  \caption{Architecture of Public Blockchain}
   \label{fig:arch}
\end{figure}

This layered architecture, illustrated in Figure \ref{fig:arch}, describes how the data is stored (Data Layer) and shared (Network Layer) between different participants of the network. Once the data is shared with peers in the network, the network is tasked with agreeing a single view of the data (Consensus Layer). Public blockchains attain consensus in the network by incentivizing non-malicious participants using an incentive mechanism (Incentive Layer). Incentive and consensus operations are performed by the execution of computational scripts (Contract Layer). The computational capabilities of a blockchain are not just limited to these two operations; many different applications can be built on top of the blockchain such as cryptocurrencies and decentralized applications (DAPPS) (Application Layer) \cite{antonopoulos2018mastering}.

In the following subsection, we describe these layers in-depth:

\subsubsection{Data Layer\label{2.1.1background}}
 The data layer contains the definition of the data structure used by the system, including how transactions are stored, thus encompassing the transactions component proposed by Bonneau et al. (2015) \cite{bonneau2015sok}. Other data layer components include the cryptographic primitives employed on the blockchain. The network participants must adhere to the data layer specifications to participate in the network, i.e., use the same protocol to communicate. Application layer blockchain clients implement these specifications for the end-user. 

\subsubsection{Network Layer\label{2.1.2background}}
The network layer specifies the behavior of the nodes (network participants) in a distributed network. This behavior includes the network connection establishment and intercommunication mechanism. The network layer is responsible for the discovery of other nodes on the network and for efficient communication among nodes. The network layer serves as the information dissemination mechanism of the system. This network layer is identical to the network subsystem in the structure proposed by Judmayer et al. (2017) \cite{judmayer2017blocks}.

\subsubsection{Consensus Layer\label{2.1.3background}}
Once the participating nodes are connected in a predefined topology, the next step is to generate blocks to contribute to the growing ledger. As all the participating nodes are tasked with the creation of the next block, it is crucial that the network can agree on a single state of the ledger. The aim of the blockchain network is to deterministically agree on a single state of the data. The consensus layer assures that the network reaches a consensus with a certain degree of assurance. 

\subsubsection{Incentive Layer\label{2.1.4background}}

This deterministic assurance is based on the assumption of an honest majority i.e. the network has at least higher than 50\% non-malicious participants. Blockchain systems use incentive engineering to ensure that the majority of the network is honest \cite{SaiBuckleyLeGear2019}. This incentive is often in the form of a block reward which is assigned to the node that successfully adds a new block to the blockchain. The incentive layer describes the mechanism used for issuance of reward and the distribution of reward. This layer acts as an interface between the user-facing layers and the technical implementation layers.

\subsubsection{Contract Layer\label{2.1.5background}}
To process transactions in the network, Bitcoin uses a scripting language called \textit{script} \cite{antonopoulos2017mastering}. This scripting language is significantly limited in terms of functionality as it lacks Turing completeness \cite{buterin2013ethereum}. One example of this is the lack of loops in \textit{Script}. Despite the lack of such functionality, the scripting language serves as the building block of Bitcoin cryptocurrency, enabling complex financial transaction processing. 

The limitations on the scripting language of Bitcoin served as a motivation for Ethereum's developers \cite{wood2014ethereum}. Ethereum implements a Turing complete computing engine on top of a distributed blockchain. Applications on top of the blockchain exploit this programmable nature of blockchain.

\subsubsection{Application Layer\label{2.1.6background}}
Public blockchains provide a mechanism that can be used to interact with and run user-defined code on the computing engine provided by the contract layer. JSON Http API is an example of one such public API provided by Ethereum \cite{lee2019using}. These public APIs serve as an interface between different Broker-Dealer services such as Wallets and Exchanges and the blockchain. These services are primarily used by end-users to interact with the blockchain \cite{chu2018broker}.

\section{Methodology\label{3methodology}}

\begin{figure}[h]
  \centering
  \includegraphics[width=\linewidth]{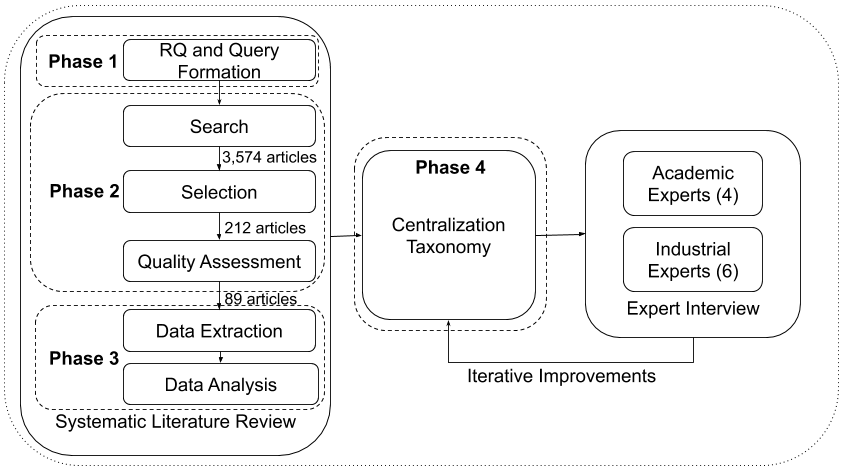}
  \caption{Methodology}
   \label{fig:mehodology}
\end{figure}

In this section, we describe the research methodology employed for our systematic literature review (SLR) of blockchain through which we sought to provide a more cohesive overview of centralization in public blockchains. We follow the SLR guidelines proposed by Kitchenham et al. (2004) \cite{kitchenham2004procedures} to identify the factors associated with centralization. We then use a classification scheme based on the generic architecture presented in Section \ref{2.2background} to map the identified factors and associated measurement techniques. This mapping is loosely based on the approach proposed by Petersen et al. (2008) \cite{petersen2008systematic}. The mapping of obtained data to the generic architecture produces an initial taxonomy, which we then refined by conducting ten expert interviews to improve the taxonomy. This process is graphically illustrated in Figure \ref{fig:mehodology}.

\subsection{Systematic Literature Review\label{3.1methodology}}
\begin{figure}[h]
  \centering
  \includegraphics[width=\linewidth]{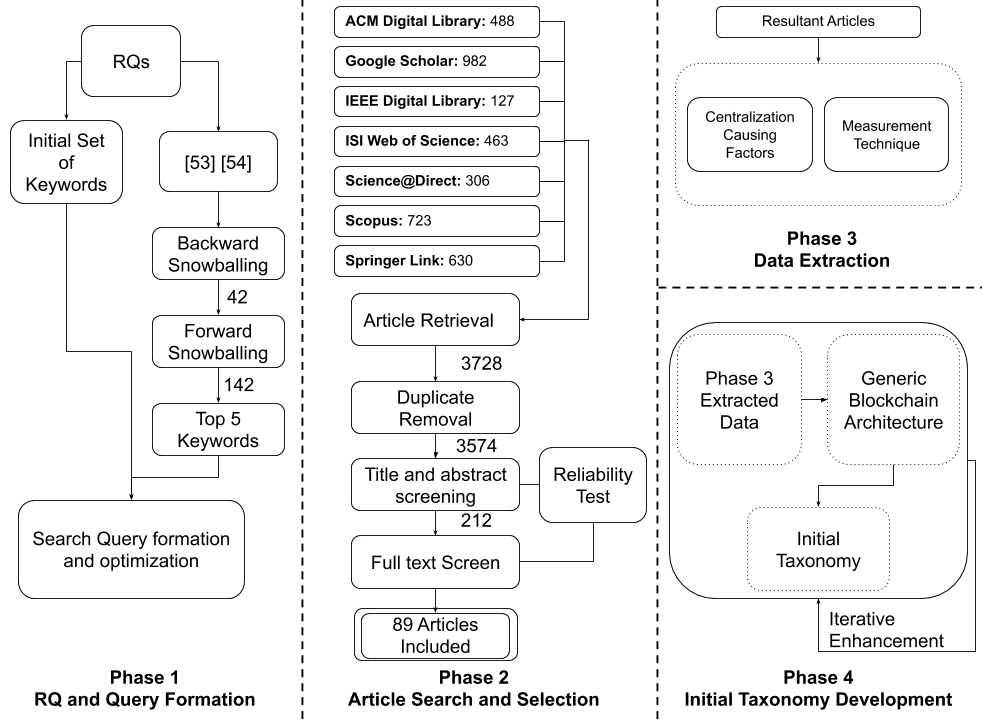}
  \caption{Overview of Systematic Literature Review}
   \label{fig:slr}
\end{figure}

The systematic literature review guidelines suggested by Kitchenham et al. (2004) \cite{kitchenham2004procedures} span four phases: 
\begin{itemize}
\item In the first phase, we define the two primary research questions for the review and produce relevant keywords for the subsequent search.
\item  In phase two, we systematically extract relevant articles from leading research repositories. We filter the resultant articles through a manual review of titles and abstracts.
\item In phase three, the shortlisted articles are then used for data extraction, which is driven by an extraction protocol.
\item  In phase four, we perform the mapping of the data extracted from phase three to the generic architecture presented in Section \ref{2.2background},leading towards an initial taxonomy of centralization in public blockchains. 
\end{itemize}
 Figure \ref{fig:slr} illustrates the literature review employed in the study in more detail.

\subsubsection{Phase 1: Research Questions and Query Formation\label{3.1.1methodology}}

The primary aim of our review is to provide richer insight into the different types of centralization present in public blockchain. We also identify techniques used to measure these aspects of centralization quantifiably. This will inform the development of our initial centralization taxonomy of public blockchains. We define the research questions of our study as follows:

\begin{itemize}
\item \textbf{RQ1:} What are the different aspects of centralization in public blockchains?
\item \textbf{RQ2:} What techniques are employed to measure these centralization aspects?
\end{itemize}

Regarding RQ1, if a paper presented a novel centralization-causing factor, it is mapped to the architecture. If our generic architecture cannot accommodate the identified factor, we modify the architecture. This process is repeated for every novel factor identified. If a paper identified a factor already present in our taxonomy, we retain the reference to the article, using number-of-articles to define a proxy for the significance of that particular factor. 

For every identified factor, we also recorded any measurement technique used to quantify the factor. If multiple papers employ different measurement techniques for a single factor, we retained all measurement techniques.

These research questions form the basis of article identification and selection, as they define the relevance of a particular article to our review. As we aim to capture factors from different socio-technical aspects of the blockchain, we conducted an exhaustive search on the following leading digital repositories: \textbf{Google Scholar, ACM Digital Library, IEEE Digital Library, ISI Web of Science, Science Direct, Scopus and Springer Link}. These repositories provided us with access to a wealth of articles, including gray literature. 

Having identified the search repositories, we formed the search query. We adopted a systematic approach to keyword generation to form the search query:

\begin{enumerate}
\item \textbf{Initial set of keywords}: We formulated an initial set of keywords for the search consisting of ``Blockchain" and ``Centralization" with the following synonyms and alternate words: \\ 
\textbf{Blockchain:} \textit{bitcoin, ethereum, blockchain, cryptocurrencies, cryptocurrency, distributed ledger, DLT, Merkel tree, smart contract platform, tokenized asset}. \\
\textbf{Centralization:} \textit{centralisation, centralism, consolidation, decentralisation, decentralization, devolution, dominating, domination, managed, monopolisation, monopolization, monopoly, singular, unipolar}.

\item \textbf{Text Corpus Creation}: Complementary to the initial set of keywords, we also reviewed existing studies on centralization to extract more relevant keywords. We selected the two most cited relevant studies from Google Scholar \cite{gencer2018decentralization,gervais2014bitcoin}. We performed forward, and backward snowballing on these two articles and generated a list of the most used keywords from this set. We selected the top 5 keywords from this set. This leads to the inclusion of \textit{``digital currency"} and \textit{``oligopoly"} to our initial set of keywords.  
\end{enumerate}

The resultant queries from query formation step are present in Appendix.

\subsubsection{Phase 2: Article Search and Selection \label{3.1.2methodology}}

Given that decentralization is fundamental to a public blockchain, we expect that the search will return a high number of articles. We implement a filtering process to limit the search to relevant articles. We restrict our search to articles published in English after the introduction of Bitcoin in 2009. We refrain from treating citations as a proxy for quality to filter articles, as it has been questioned in the past \cite{galster2013variability}.

After the execution of a search query, Google Scholar returned the highest number of articles with 4,380 results. However, due to the restrictions imposed by Google Scholar, we can only retrieve the first 1000 most relevant articles \cite{razzaq2018state}. After applying the language and publication date constraints, we retrieved 982 articles from Google Scholar. We also retrieved additional 2737 articles from all other sources resulting in a total of 3728 articles. All of these articles were cross-checked to identify duplicate entries. After the removal of duplicate articles, the final set contained 3572 articles \footnote{A list of selected articles is available at www.github.com/ashishrsai/centralization}. 

Due to the high number of articles, we first analyzed the title and abstract to establish relevance. This was based on explicit inclusion criteria. The shortlisted, relevant articles were then scanned further to assign a quality score. These shortlisted articles were assessed for quality with regards to our research questions. To ensure that the assessment process is reliable, we followed the inclusion criteria for titling, abstraction, and full-text screening. This process obeyed the following inclusion criteria:
\begin{enumerate}
\item The paper's title mentions centralization, or any of the synonyms mentioned above, or is potentially relevant to the study of centralization.
\item The abstract is relevant to the identification or measurement of centralization-causing factors. 
\end{enumerate}
During the review of the title, we tried to avoid eliminating articles that might have some relevance to the topic of centralization. This relevance was evaluated by the review of the abstract. We excluded articles that did not pass both criteria.

The first author conducted this analysis. To test for reliability, we performed cross-validation by following Fleiss  al. (1973) \cite{fleiss1973equivalence}. We specifically use the guidelines proposed by Sim  al. (2005) \cite{sim2005kappa} for the calculation of sample size. We select 89 articles with a confidence level of \textbf{95\%} and a margin of error of \textbf{10\%}. This sampling contained an equal number of accepted and rejected articles by the first author to eliminate the possibility of only sampling accepted or rejected articles. The second author was then tasked with the evaluation of these 89 articles based on the guidelines provided above. Results from the cross-validation suggest that both the reviewers were in almost perfect agreement over the acceptance and rejection of the articles with the Cohen's Kappa\footnote{Cohen's kappa is a statistic measure of the agreement between two raters based on the classification of items in mutually exclusive categories.} exceeding 0.8 \cite{landis1977measurement}. 

Using this process, we retrieved 212 relevant articles for our study. Subsequently, we performed quality assessment of these articles by conducting full-text review. We assigned a quality score between 0 to 2 based on the relevance of the article to our research question. Table \ref{tab:qualitymatrix} outlines the assignment matrix employed for quality assessment.

\begin{table*}
\centering
  \caption{Quality Assignment Matrix}
  \label{tab:qualitymatrix}
  \begin{tabular}{|c|c|c|}
    \hline
    Attribute & No & Yes \\
    \hline
    1. Centralization Factor Identified & 0.0 & 1.0  \\ \hline
    2. Factor Measurement Technique Proposed & 0.0 & 1.0  \\ \hline
  \end{tabular}
\end{table*}

We reviewed each article on two attributes - 1) factor identification and 2) measurement techniques used. If an article identifies a novel centralization-causing factor, we assign a score of 1.0 for Attribute 1. Articles that do not identify a novel centralization or refer to already identified factors are assigned a score of 0.0 for Attribute 1 \footnote{Although we do record the paper, as this helps us identify the significance of that centralization aspect in the literature.}. 

We follow a similar quality assignment scheme for Attribute 2, where we assign a score of 1 for the identification of a novel measurement technique. Articles not proposing or using any existing measurement techniques are assigned a score of 0.0 for attribute 2.

To ensure that the quality assignment process is reliable, we again perform a similar reliability test but with a smaller data set of 9 articles. We observe that both the reviewers (first and fourth authors) agree on eight score assignments with one score difference for the ninth article. This disagreement is resolved when the article is reviewed by the third author.

This filtering process resulted in a set of 89 articles. These articles are used in the third phase of our study: Data Extraction.

\subsubsection{Phase 3: Data Extraction\label{3.1.3methodology}}
Having identified relevant studies, the next step is to extract relevant data from them. For this purpose, we design a protocol to analyze the articles towards the development of an initial taxonomy of centralization. In this context, we focused on the factors identified and measurement techniques proposed or used. We reviewed all of the shortlisted articles to create a list of factors and associated measurement techniques. The extracted data from this step serves as a building block for our taxonomy.

\subsubsection{Phase 4: Development of Initial Taxonomy \label{3.1.4methodology}}
\begin{figure}[h]
  \centering
  \includegraphics[width=\linewidth]{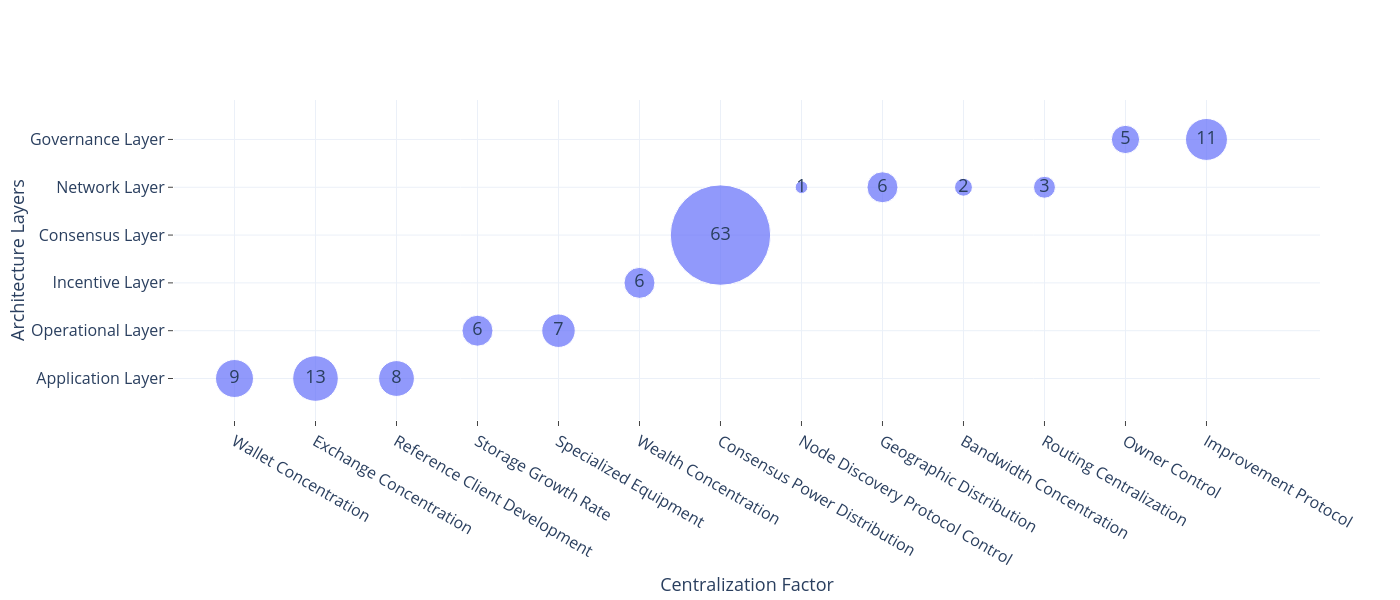}
  \caption{Article Titling and Abstraction Process}
   \label{fig:bubble}
\end{figure}

As we aim to structure the findings of the review in an initial taxonomy, we use the data extracted in Phase 3 and map it to appropriate layers in the generic blockchain architecture. We repeat this process for all identified factors; if a factor cannot reasonably be mapped to the existing layers, we typically refine the architecture by including an additional layer. This iterative refinement results in a blockchain architecture specific to the study of centralization. Results from this mapping analysis are illustrated in Figure \ref{fig:bubble}. Out of all shortlisted articles, 63 considered the consensus layer as prone to centralization, the highest reported count for any layer in our survey: This is represented in Figure \ref{fig:bubble} by the size of the bubble, but we discuss these results in more depth in Section \ref{4taxonomy}.

To further validate the initial taxonomy and refined architecture, we conducted interviews with industry and academic experts. 

\subsection{Interview with experts\label{3.2methodology}}

The initial taxonomy, as referred to in Section \ref{3.1methodology}, is based on the review of existing literature. To raise confidence that the initial taxonomy proposed by the study provides relevant coverage and is accurate, we further refine and validate it by interviewing experts. 

To identify experts in the blockchain field, we relied on the epicenters of the bibliographic map generated by \cite{ramona2019bitcoin}. We approached 112 active researchers based on their prominence determined by their location on the bibliographic map. Out of 112 researchers approached for the study, we received a response from 10 and subsequently interviewed them. We interviewed four academic experts ($I_1$ to $I_4$) and six experts from industry ($I_5$ to $I_{10}$).
Interviews were typically one hour in duration and involved open-ended questions \footnote{The interview script is available at www.github.com/ashishrsai/centralization.}. These open-ended questions were designed to: 
\begin{enumerate}
\item Extract the view of the expert on centralization and the significance of it in their respective field, i.e., security, economics, information systems, and industrial application.
\item If needed, refine the taxonomy and/or the architecture. 
\item Validate the generic architecture of the blockchain used in this study (Section 2).
\item Assess the accuracy of the initial centralization taxonomy. 

\end{enumerate}

The transcripts of these interviews are available in anonymized form \footnote{The transcripts can be obtained from www.github.com/ashishrsai/centralization.}. These transcripts are color-coded based on the relevance of the conversation to factor identification and measurement \footnote{More details on the coding scheme provided in Appendix.}. 

\section{Taxonomy of Centralization of Public Blockchain\label{4taxonomy}}
In this Section, we map the results of the systematic review, and the interviews with experts, to the initial taxonomy of centralization outlined in Table \ref{tab:taxonomy}.

As discussed in Section \ref{3.1methodology}, this generic architecture is refined to reflect the centralization-related aspects of the blockchain better. To this end, we refined the generic architecture by removing the Data and Contract layers as none of the surveyed articles suggested any centralization aspects for either of these layers.
As can be seen from Table \ref{tab:taxonomy}, on average two centralization factors were identified for each resultant layer. As is also presented in the table, there are some factors for which there are no proposed measurement techniques (for example 'Wallet Concentration'). We also note that the existing generic architecture was unable to capture governance-related aspects of the blockchain system. For example, as blockchain systems evolve, it is crucial to have a mechanism to handle improvements such as security patches of the system. We account for the governance-related aspects of centralization by including a Governance Layer.

Another set of centralization causing issues that the generic architecture does not capture are associated with the operation of a node on the network. These issues include the computational requirements for participation, such as proprietary hardware and storage. In accordance with the recommendation of interviewee $I_{10}$, we include an Operational layer to represent the centralization associated with operating as a node on the blockchain.

\begin{table*}
\centering
  \caption{Taxonomy of Centralization in Public Blockchains}
  \label{tab:taxonomy}
  \begin{tabular}{|c|c|p{4cm}|}
    \hline
    Layer & Centralization Factor & Measurement Techniques \\
    \hline
  Application Layer & Wallet Concentration &   Not Found \\
                        & Exchange Concentration & Centrality \& Percentage Value \\
                        & Reference Client Concentration & Satoshi Index \\
                        \hline
    Operational Layer & Storage Constraint & Ratio of Growth \\
                    & Specialized Equipment Concentration & Not Found \\
                        \hline
   Incentive Layer & Wealth Concentration & Gini Coefficient \& Percentage Value\\
                     \hline
    Consensus Layer & Consensus Power Distribution & Percentage Value \& Gini Coefficient \& Theil Index \& Centralization Factor\\
    \hline
    Network Layer & Node Discovery Protocol Control & Not Found \\
                     & Geographic Distribution & Gini Coefficient \& Latency\\
                     & Bandwidth Concentration & Clustering of Provisioned Bandwidth \\
                     & Routing Centralization & AS-Level Coverage \\
                     \hline
    Governance Layer & Owner Control & Fractional Measurement \\
                      & Improvement Protocol & Centrality metrics \\
                      
   \hline
  \end{tabular}
\end{table*}

Table \ref{tab:taxonomyWithCitation} considers the factors identified in Table \ref{tab:taxonomy} from the perspective of 'prevalence-of-occurrence' in the literature and the interviews, where  prevalence is considered as a proxy for whether the factor is ``established'' or not.  The literature references in the table identify that particular factor as a potential source of centralization\footnote{A complete list of articles is available in Appendix.}. The interviewer identifiers are used to indicate explicit recognition of the factor as a contributor to centralization in the associated interview. Interestingly, based on the data presented in this table, most of the factors can be considered well established, with the possible exception of Bandwidth Concentration and Routing Centralization. Even though Node Discovery Protocol Control was only referred to by one academic article, the majority of interviewees perceived it as a relevant factor. 

Based on our taxonomy, we define centralization of public Blockchains as \textit{the process by which one or more architectural dimensions (aspects) of the Blockchain are restrictive to the majority of participants by direct or indirect economic, social, or technical constraints}. We report a total of 13 aspects spread over six architectural layers. The governance layer aims to capture the social constructs of building and maintaining a public blockchain, specifically reporting on the incentives to build (Owner Control) and maintain a public blockchain (Improvement Protocol). The governance layer feeds into the economic aspects of the Blockchain in forms of incentives, this is captured by the Incentive layer, where we review the wealth inequality (Wealth Concentration). This inequality is in part caused by the technical constraints of participation ranging from Networking aspects such as bandwidth and routing requirements to operational requirements such as storage and specialized pieces of equipment for participation. These higher storage and specialized equipment requirements restrict participation in the consensus, which is observable in the consensus layer. We also report on the centralization of end-user applications such as wallets and exchanges. The following subsections discuss the taxonomy in detail.

\begin{table*}
\centering
  \caption{Centralization Causing Factors Found in Literature and Interviews}
  \label{tab:taxonomyWithCitation}
  \begin{tabular}{|c|p{4.5cm}|l|}
    \hline
    Centralization Factor & Refereed Articles & Interviews \\
    \hline
   Wallet Concentration & R11, R13, R36, R40, R76, R78, R84, R86, R88 & $I_4$,$I_5$,$I_7$,$I_8$\\ \hline
    Exchange Concentration & R11, R13, R27, R34, R37, R40, R57, R64, R73, R78, R84,
R86, R89 & $I_1$,$I_3$,$I_4$,$I_5$,$I_7$,$I_8$ \\ \hline
    Reference Client Concentration & R4, R6, R8, R26, R36, R50, R67, R83 & $I_2$,$I_5$,$I_8$,$I_9$,$I_{10}$ \\ \hline
                       
   Storage Growth Rate & R9, R24, R38, R39, R63, R80 & $I_2$,$I_{10}$ \\ \hline
   Specialized Equipment Concentration &  R23, R51–R53, R55, R62, R67
 & $I_4$,$I_5$,$I_7$,$I_8$,$I_9$,$I_{10}$\\ \hline                     
   Wealth Concentration & R16, R51, R52, R55, R62, R67 & $I_1$,$I_2$,$I_3$,$I_4$,$I_5$,$I_6$,$I_7$,$I_9$ \\ \hline
                    
    Consensus Power Distribution & R1–R3, R5, R7, R9, R11–R17, R19–R22, R25, R26, R28–
R33, R35, R36, R39, R40, R42–R47, R49, R52–R56, R58,
R60, R61, R65–R72, R74–R79, R81, R82, R87 & $I_1$,$I_2$,$I_3$,$I_4$,$I_5$,$I_6$,$I_7$,$I_8$,$I_9$,$I_{10}$\\ \hline
    
    Node Discovery Protocol Control & R59  & $I_1$,$I_2$,$I_3$,$I_5$,$I_{10}$ \\ \hline
     Geographic Distribution &   R5, R30, R40, R47, R50, R76  & $I_1$,$I_2$,$I_3$,$I_4$,$I_5$,$I_6$,$I_7$\\ \hline
    Bandwidth Concentration &  R35, R87 & $I_2$,$I_{10}$ \\ \hline
    Routing Centralization & R3, R20, R35 & $I_2$\\ \hline
    Owner Control & R14, R18, R26, R41, R48 & $I_1$,$I_4$,$I_5$,$I_7$,$I_8$,$I_9$ \\ \hline
    Improvement Protocol & R4–R6, R10, R26, R36, R41, R48, R76, R83, R85 & $I_1$,$I_2$,$I_3$,$I_4$,$I_5$,$I_7$ \\ \hline
                      
  \end{tabular}
\end{table*}

\subsection{Governance}

Blockchain, like any other information system, is subject to evolutionary changes that are governed by a governance structure. These evolutionary changes may include security patches, scalability provisions, and improvement proposals. Wang  al. (2017) \cite{wang2017internal} theorizes the relationship between the value proposition of blockchain and the governance structure in place. They reason that the core value proposition of blockchain is rooted in decentralization. This property of decentralization is considered valuable by investors. 

Decentralized governance was also indicted as a vital component of public blockchains by our interview participants. 80\% mentioned governance as a significant centralization threat ($I_1$,$I_2$,$I_3$,$I_4$,$I_5$,$I_7$,$I_8$,$I_9$). This is best illustrated by a quote from $I_1$, with respect to the implication of centralized governance structure: \textit{``if you are talking about the centralization of governance, that for me is the prime example of a private permissioned Blockchain"}.

Despite the significance of decentralization for blockchain, Wang  al. (2017) \cite{wang2017internal} argue that a high level of decentralization may slow down the strategic decision-making process. Contrary to the proposition in favor of some centralization by Wang  al. (2017) \cite{wang2017internal}, Gervais  al. (2014) \cite{gervais2014bitcoin} argue against the concentration of decision making power by pointing out instances of unilateral decision making by core developers in the short history of bitcoin; for example, when the core developers unilaterally decided to lower the minimum transaction fee. This criticism of governance centralization is shared by Roubini (2018) \cite{122} who criticizes the centrality of control over governance as it may concentrate the decision power to a few entities involved in governance of the blockchain. Atzori (2015) \cite{87} expands the analysis of blockchain governance issues towards the emergence of blockchain governance oligarchy. Azouvi  al. (2018) \cite{azouvi2018egalitarian} conducts an empirical analysis of two of the most prominent blockchain projects, Bitcoin and Ethereum, by comparing the state of governance to other major open-source projects. They conclude that control governance is usually concentrated in a handful of people in Bitcoin and Ethereum, which is a big centralization factor.

As reported by Wang et al. (2017) \cite{wang2017internal}, the centralization on the governance layer may not be detrimental due to the advantages of rapid strategic decision-making. We expand on the argument in favor of some centralization \cite{wang2017internal} in Section \ref{discussion}, where we discuss how the adverse impact of centralization varies across the different layers of the taxonomy. 
 
Based on the literature review and subsequent interviews, we further divide the issue of governance into \textit{owner control} and \textit{improvement protocol}. These results are presented in Table \ref{tab:govTable}. 

\begin{table}
\small
    \centering
    \caption{Categories of centralization in Governance Layer}
    \label{tab:govTable}
  \begin{tabular}{|l|S|S|S|S|}
    \hline
    \multirow{2}{*}{Ref} &
      \multicolumn{2}{c|}{Owner Control} &
      \multicolumn{2}{|c|}{Improvement Protocol}  \\ \cline{2-5}
      & {Identification} & {Measurement} & {Identification} & {Measurement}  \\\hline

    R4 & {$\times$} & {$\times$} & {\checkmark} & {$\times$} \\
    R5 & {$\times$} & {$\times$} & {\checkmark} & {$\times$} \\
    R6 & {$\times$} & {$\times$} & {\checkmark} & {Centrality Metrics} \\
    R10 & {$\times$} & {$\times$} & {\checkmark} & {$\times$} \\
    R14 & {\checkmark} & {$\times$} & {$\times$} & {$\times$} \\
    R18 & {\checkmark} & {Fractional Measurement} & {$\times$} & {$\times$} \\
    R26 & {\checkmark} & {$\times$} & {\checkmark} & {$\times$} \\
    R36& {$\times$} & {$\times$} & {\checkmark} & {$\times$} \\
    R41& {\checkmark} & {$\times$} & {\checkmark} & {$\times$} \\
    R48& {\checkmark} & {$\times$} & {\checkmark} & {$\times$} \\
    R76& {$\times$} & {$\times$} & {\checkmark} & {$\times$} \\
    R83& {$\times$} & {$\times$} & {\checkmark} & {$\times$} \\
    R85& {$\times$} & {$\times$} & {\checkmark} & {$\times$} \\
    \bottomrule
  \end{tabular}
   
\end{table}

  \subsubsection{Owner Control} As described by Wang et al. (2017) \cite{wang2017internal}, the developers of the blockchain often retain some control over the implementation on the governance level. This can be in the form of, for example, the native cryptocurrency owned by the developers. Wang et al. (2017) \cite{wang2017internal} describes this as \textit{Owner Control}. 
  \\ \textbf{Measurement Technique }: This type of owner control can be measured by examining the total cryptocurrency accumulated by the owners in the early adoption period \cite{wolfson2015bitcoin}. This early adoption period also includes the pre-mined \footnote{Pre-mined cryptocurrency refers to the native cryptocurrency issued with the creation of the first block in the blockchain.} cryptocurrency \cite{wang2017internal}. We report  studies such as \cite{wolfson2015bitcoin,chohan2019cryptocurrencies} that have implemented a proportional measure to quantify owner control. Owner control can be measured as the fraction of the total allowed cryptocurrency if the supply is capped, as measured by Equation 1, where $C_{OwnerControl}$ represents the fraction of total cryptocurrency that the owner controls. 
  
\begin{equation}
    C_{OwnerControl} = V_{OwnerBalance}/V_{CappedSupply} 
\end{equation}

If the supply is uncapped, owner control is measured as the fraction of total currency in circulation, as illustrated in Equation 2.

\begin{equation}
    C_{OwnerControl} = V_{OwnerBalance}/V_{CurrentSupply} 
\end{equation}
  
Most interview participants indicated that the use of fractional measurement for owner control was appropriate. However, $I_9$ suggested a refinement:  \textit{``The fractional calculation of the owner control varies with the supply; a simpler approach might be to use a metric such as how much power over the network can be achieved with the money in the owner control. How much hardware can you afford, and what hash power can you get with it. Relating the cryptocurrency to the hashing power would be more informative"}.
  
 \textbf{Implication of high owner control} : Depending on the consensus mechanism used, the owner control has severe impacts on the network. This adverse impact is particularly worrying in the case of Proof-of-stake based cryptocurrency, where the consensus power is determined by the quantity of native cryptocurrency owned by the participant. Having a large amount of pre-mined or early adoption period accumulated cryptocurrency will give the owner a significant advantage over others, resulting in a more centralized network. This high consensus power pose a security threat as an owner with over 50\% consensus power can conduct a double spending attacks. Ethereum is a prime example of such wealth concentration due to pre-mined cryptocurrency. 
 
 The Ethereum platform was crowdfunded by investors who were rewarded in the form of ETH\footnote{ETH is the ticker mark for Ether, the cryptocurrency used by Ethereum platform.} during the creation of the first block in Ethereum. An estimated 60 Million ETH were distributed among the early investors; another 12 Million were distributed among the developers of Ethereum \cite{ethereumtransactioninformation}. We calculate the value of $C_{OwnerControl}$ by considering the 12 Million pre-mined ETH that developers control and the total current supply of ETH obtained (from \cite{totalETH}): 
\begin{equation}
    C_{OwnerControl} = 12,000,000 /106,514,407.78  = 0.11
\end{equation}
 
It should be noted that the value of $C_{OwnerControl}$ feeds into the issue of Wealth Concentration, which is a significant cause of economic centralization. A high wealth concentration in a cryptocurrency is against the founding principle and premise of cryptocurrency providing a more even monetary system. This can consequently disincentivize the adoption.

 \subsubsection{Improvement Protocol} As discussed earlier, evolutionary changes require blockchains to have a robust governance structure in place. As decentralized blockchains do not have any authorized entities moderating the changes, the process of moderation is delegated to the participants. Bitcoin improvement protocol (BIP) is a prime example of such an improvement system \cite{anceaume2016safety}. The formal voting protocol, such as that in BIP, is used to establish consensus over proposed changes, often through voting. 60\% of interview participants ($I_1$,$I_2$,$I_3$,$I_4$,$I_5$,$I_7$) mentioned that the improvement protocol performs an essential function in the network with $I_7$ suggesting: \textit{``Whoever controls the improvements will inevitably shape the future of the network"}.
 
 The literature review points out the similarities between the Python Enhancement Proposals and BIPs, both of which heavily draw from the \textit{``canonical"} approach to consensus \cite{de2016invisible}. In the \textit{``canonical"} based BIP, all the suggested changes have to be made available to the public for open discussion. However, the final decision as to how proposed changes will be implemented is taken by the core developers \cite{gervais2014bitcoin}. 
 
 \textbf{Measurement Technique}: The centralization in a formal voting protocol is measured by analyzing the moderation control. If specific developers or owners moderate the voting, the moderation may jeopardize the changes that developers or owners disagree with. Thus the determination of the control level is done by examining the voting protocol in place and the controls imposed on it.
As public blockchains often have an open platform for proposing improvements, such as BIP for Bitcoin, and EIP for Ethereum, Azouvi et al. (2018) \cite{azouvi2018egalitarian} suggests reviewing the number of improvement proposals made by each author and the respective states of those proposals (i.e., approved, rejected or under review). The authors also suggest reviewing the comments on each proposal to examine the discussions. Based on the data obtained from the author and number of proposals complemented by comments per author on the proposal, Azouvi et al. (2018) \cite{azouvi2018egalitarian} suggests calculating centrality metrics for the centralization measurement. 

These centrality metrics include Mean, Median, interquartile range (IQR), and interquartile mean (IQMean). IQR is a measure of variability that assists in locating where the majority of values lie in the data sample. It is calculated as the difference between $75^{th}$ and $25^{th}$ percentiles of the data. However, IQR is sensitive to noisy outliers, which can impact the overall result. This can be overcome by using the IQMean, which allows us to eliminate the outliers from our data set by calculating the median of IQR.

\textbf{Implication of control over improvement protocol}: If a subset of all participants moderate the improvement protocol, it will result in control over improvements or modifications to the network. The debate over block size in Bitcoin an example of an issue arising due to this type of control over the network \cite{bitcoin_2019,de2016invisible}. Other significant control implications over the improvement protocol include the unilateral decision making in both Bitcoin and Ethereum, where the governance structure implemented a change not widely supported by the community. This includes the notable transaction fee reduction in Bitcoin \cite{gervais2014bitcoin} and Ethereum hard fork due to DAO attack which led to the subsequent creation of Ethereum classic \cite{wirdum2016rejecting}. More incidents of unilateral decision making include the changes to the Ethereum consensus algorithm in 2018, where developers decided to modify the algorithm to disable newer mining hardware \cite{kim2019ethics}. These incidents not only represent the lack of a systematic governance model in terms of improvement but also present a challenge in terms of newer participation and updates. This type of centralization impacts the presumed open nature of the Blockchain, which is one of the core contributions of Blockchain to the field of financial technologies.

\subsection{Network} 
The network layer acts as the information dissemination mechanism for the blockchain instance. As the decentralized network cannot have centralized nodes that act as relay points to transmit messages between the participants, the network is largely a peer-to-peer system. The network layer acts as the information dissemination mechanism for the blockchain instance. As the decentralized network cannot have centralized nodes that act as relay points to transmit messages between the participants, the network is largely a peer-to-peer system. This peer-to-peer network serves as an essential security and usability measure as pointed out by $I_8$:\textit{ ``In this peer to peer network, there is no single point of failure and participants can join and leave the network without risking interruption or degradation of the network"}.

Network connectivity of a node is an important aspect of performing the mining operation \cite{sapirshtein2016optimal}. Higher network connectivity results in a higher likelihood of adding the next block on the longest chain as the miner can propagate the block to a large number of nodes in the network. This interplay between the reward from adding a block to the blockchain and network connectivity has resulted in networking phenomena such as strategizing networking resource concentration in the form of bandwidth \cite{gencer2018decentralization} and strategizing geographic distribution of nodes in the network \cite{58,roubini_2018}.

Based on the literature review, we identify another source of centralization on the network layer as the topology formation of the network. This formation includes the node discovery protocol for finding peers in the network \cite{36} and the routing structure of the network \cite{apostolaki2017hijacking}. The relevant studies identified by our review are presented in Table \ref{tab:cenNet}. We describe each of the outlined factors in detail in the following Subsections.

\begin{table}
\tiny 
    \centering
    \caption{Categories of centralization in Network Layer}
    \label{tab:cenNet}
  \begin{tabular}{|l|S|S|S|S|S|S|S|S|}
    \hline
    \multirow{2}{*}{Ref} &
      \multicolumn{2}{c|}{Node Discovery} &
      \multicolumn{2}{c|}{Geographic Distribution}  &
      \multicolumn{2}{c|}{Bandwidth} &
      \multicolumn{2}{c|}{Routing}  \\ \cline{2-9}
      & {Identification} & {Measurement} & {Identification} & {Measurement}  & {Identification} & {Measurement} & {Identification} & {Measurement}  \\
      \hline
    R3 & {$\times$} & {$\times$} & {$\times$} & {$\times$} & {$\times$} & {$\times$} & {\checkmark} & {AS Coverage} \\
    R5 & {$\times$} & {$\times$} & {\checkmark} & {$\times$} & {$\times$} & {$\times$} & {$\times$} & {$\times$}\\
    R20 & {$\times$} & {$\times$} & {$\times$}& {$\times$} & {$\times$} & {$\times$} & {\checkmark} & {$\times$}\\
    R30 & {$\times$} & {$\times$} & {\checkmark} & {$\times$} & {$\times$} & {$\times$} & {$\times$}& {$\times$}\\
    R35 & {$\times$} & {$\times$} & {\checkmark} & {Latency} & {\checkmark} & {Clustering} & {\checkmark} & {AS Coverage}\\
    R40 & {$\times$} & {$\times$} & {\checkmark} & {$\times$} & {$\times$} & {$\times$} & {$\times$} & {$\times$}\\
    R47 & {$\times$} & {$\times$} & {\checkmark} & {$\times$} & {$\times$} & {$\times$} & {$\times$} & {$\times$}\\
    R50& {$\times$} & {$\times$} & {\checkmark} & {$\times$} & {$\times$} & {$\times$} & {$\times$} & {$\times$}\\
    R59& {\checkmark} & {$\times$} & {$\times$} & {$\times$} & {$\times$} & {$\times$} & {$\times$} & {$\times$}\\
    R76& {$\times$} & {$\times$} & {\checkmark} & {$\times$} & {$\times$} & {$\times$} & {$\times$} & {$\times$}\\
    R87& {$\times$} & {$\times$} & {$\times$} & {$\times$} & {\checkmark} & {$\times$} & {$\times$} & {$\times$}\\
    \hline
  \end{tabular}
   
\end{table}

  \subsubsection{Node Discovery Protocol Control} In a peer-to-peer topology, participating nodes directly communicate with other participants to transmit data packets. A node discovery protocol is used to discover nodes in the network with which to communicate \cite{miller2015discovering}. The node discovery protocol often relies on a set of seed DNS nodes that distribute the address of other active nodes on the network. These predefined DNS nodes may be a potential source of security threat, as demonstrated by Tapsell et al. (2018) \cite{tapsell2018evaluation} and Jin et al. (2017) \cite{jin2017blockndn}. If one of the seed nodes becomes inaccessible, it may result in many participants of the network becoming undiscoverable. As the new nodes in the network discover others by querying these predefined seed DNS nodes, the literature identifies seed nodes as a contributor to centralization on the network layer \cite{36}. 
  
  \textbf{Measurement Technique }: After the review of all relevant articles in our study, we conclude that no measurement technique focuses on the Node Discovery protocol. Studies such as \cite{jin2017blockndn,tapsell2018evaluation} investigate the issue of seed DNS nodes from a security perspective, specifically focusing on the single point of failure issue. We reason that further investigation into centralization in node discovery level is warranted due to the significant security threats that it poses.  
  
  \textbf{Implication of control over DNS}: Centralized DNS services are linked to security threats in the network \cite{jin2017blockndn}. They also allow the DNS owners to observe the participants of the network. These centralized DNS services can also act as a single point of failure, which is of particular concern in the case of a Denial of Service attack \cite{dietrich2000analyzing}. As core developers select these DNS nodes, the issue of node discovery protocol also feeds into that of trust in the core developers \cite{tapsell2018evaluation}. A malicious developer can also change the DNS seed nodes to conduct an eclipse attack. Serval Monte Carlo simulations have shown the effectiveness of such eclipse attacks on Bitcoin and Ethereum \cite{heilman2015eclipse}.  
  
  \subsubsection{Geographic distribution} Bitcoin and similar cryptocurrencies have been able to gain significant attention from governments around the world due to their decentralized uncensored nature. This has prompted many to argue that a significant concentration of the nodes in any geographic area may be a threat to the network \cite{roubini_2018}. This type of geographic concentration may lead to centralization on the network layer as the nodes become prone to geopolitical manipulation. 70\% of interview participants indicated that geographic concentration is harmful to the network. $I_6$ suggested that geographic centralisation may be disadvantageous for miners who are not centrally located: \textit{``I fear that in a geographically-focused network, people within the same geographic location will have an edge over others, they will receive and send transactions first"}.
  
  The nodes are distributed over the participating countries in the network. In an ideal case, the distribution of nodes should be equal in all participating countries so as to be able to withstand a geopolitical blockade. Findings from our review suggest there is a trend towards geographic concentration of nodes in both Bitcoin and Ethereum \cite{58,roubini_2018,54,gencer2018decentralization}.

  \textbf{Measurement Technique }: Our review suggests that the geographic location measurement in blockchain can be done by measuring latency in the peer-to-peer network \cite{58,gencer2018decentralization}. This approach draws heavily from Saroiu et al. (2001) \cite{saroiu2001measurement}, where the authors proposed using latency as a measurement tool in Gnutella. Gencer et al. (2018) \cite{gencer2018decentralization} first proposed measuring the distance between their geographically distributed nodes and other peers in the network by sending a data packet and measuring the round-trip time. Based on the round-trip time, Gencer et al. (2018) \cite{gencer2018decentralization} calculated upper and lower bounds between two remote peers in the network. If two nodes take a similar time to respond to the data packet sent by their nodes, it is reasoned that these two nodes are likely geographically close. This approach is further refined by Kim et al. (2018) \cite{58}, who consider the average of bounds for final latency estimation. 

\textbf{Implications of geographical centralization}: The most prominent issue with geographic centralization is the potential for geopolitical manipulation of the network \cite{roubini_2018}. Other issues with geographic clustering include the possibility of faster transmission of packets to nearby nodes promoting faster network propagation. This can lead to more clustering, since participant must propagate the solution to the majority of the network in order to get rewarded in Proof-of-work based blockchains. If the majority is located in a geographical cluster away from the participant, that may translate to a loss of revenue. As suggested by Gencer et al. (2018) \cite{gencer2018decentralization}, a low number of geographic clusters are considered good for the decentralization of the network. This is due to the association of potentially high block rewards due to faster network propagation. As shown in Sapirshtein et al. (2016) \cite{sapirshtein2016optimal}, network connectivity is directly related to the ability to successfully conduct selfish mining attacks, which can support a double spending attack.

\subsubsection{Bandwidth Concentration} In a public blockchain's peer-to-peer network, the network bandwidth often acts as a crucial factor in the successful propagation of data packets. In Proof-of-Work based blockchain, every consensus cycle acts as a race to first calculate the solution to the cryptographic puzzle followed by dissemination of the solution to a majority of the network. Dissemination requires a large number of network connections with peers in the network, thus increasing the bandwidth requirements. This arms race to attain higher bandwidth may lead to the centralization of mining equipment to services like a centralized data center with high bandwidth \cite{gencer2018decentralization}. 
  
  \textbf{Measurement Technique }:  Gencer et al. (2018) \cite{gencer2018decentralization} proposed measuring the bandwidth of each peer by requesting a large amount of data and estimating the speed by observing the time taken for the transmission. Once they estimate the speed of each accessible peer, they calculate and cluster the provisioned bandwidth in groups. 
  
  \textbf{Implication of bandwidth concentration}: A high bandwidth requirement may limit the participation to only the participants with significant bandwidth \cite{19}. It may also result in a high concentration of networking devices in centralized spaces such as data centers \cite{gencer2018decentralization}. This potential increment in bandwidth requirement may limit the participation to only those entities with high network capabilities making the consensus participation not viable in a domestic setting. The inability to participate in the network violates the open nature of the public blockchain preventing a widespread adoption of the technology.
  
  \subsubsection{Routing Centralization} As public blockchain networks run over the existing networking stack, they rely on the networking structure used by IP (Internet Protocol). Centralization present in the networking structure of IP transfers to the blockchain as well. Our review reports that this centralization has been studied in blockchain from the privacy \cite{feld2014analyzing} and security \cite{apostolaki2017hijacking} perspectives. Gencer at al. (2018) reports that concentration on AS-Level\footnote{Autonomous systems (AS) in computer networks refers to the collection of connected IP routing prefixes under the authority of one or more networking entities.} as a source of centralization for a public blockchain \cite{gencer2018decentralization}. Interestingly, none of the industrial participants mentioned this concern unprompted, suggesting that it might be more of an academic concern than a real-world one. However, when the concern was mentioned, one industry participant agreed.
  
  \textbf{Measurement Technique }: Our review suggests that there is a common network traversing strategy used to determine the network structure from the AS-Level perspective \cite{feld2014analyzing,apostolaki2017hijacking,gencer2018decentralization}. To measure the number of ASes in a peer to peer network, the observer node traverses the network by recursively collecting IP addresses of each peer and querying every reachable address. This process is repeated until no new reachable nodes are available in the IP list. For the determination of AS of each IP, Feld et al. (2014) \cite{feld2014analyzing} recommend using Maxmind's free Geo API\footnote{https://dev.maxmind.com/geoip/geoip2/geolite2/}. 
  
  \textbf{Implication of control over ASes}: Centralization on AS-Level is reported to have privacy implications for blockchain users as it allows more traceability on a network level \cite{feld2014analyzing}. This concentration of IP addresses under a few ASes is directly linked with potential network security issues in Bitcoin \cite{apostolaki2017hijacking} and Ethereum \cite{gencer2018decentralization}. However, these privacy and security threats remain largely academic with no real world incident reports in our sample set of articles. This is further evident through our interviews, where no academic or industrial experts pointed to control over ASes as a centralization threat unprompted.

\subsection{Consensus \label{5.3}}
 The consensus layer establishes an agreement on a single state of the data in the public blockchain. As described in Section \ref{2.1background}, in the case of Proof of Work, it is attained by inducing a race to solve a mathematical problem. The first person to solve and propagate receives a monetary reward as an incentive. The likelihood of finding the solution to the mathematical problem depends on the computational power devoted to the solution. Thus a high concentration of computational power is a direct signifier of centralization in the blockchain. As identified by articles in Table \ref{tab:consensus}, the \textbf{consensus power distribution} is a key contributor to the centralization of the Proof-of-Work based blockchain. Eight interviewees mentioned this aspect unprompted, suggesting that this is a prevalent concern. In this subsection, we review how the literature defines and measures the consensus power centralization.  

\begin{table*}
\centering
  \caption{Categories of centralization in Consensus Layer}
  \label{tab:consensus}
  \begin{tabular}{|c|p{5cm}|}
    \hline
    Consensus Power Distribution & Selected Studies \\
    \hline
    Identification & R1, R2, R3, R5, R7, R9, R11, R12, R13, R14, R15, R16, R17, R19, R20, R21, R22, R25, R26, R28, R29, R30, R31, R32, R33, R35, R36, R39, R40, R42, R43, R44, R45, R46, R47, R49, R52, R53, R54, R55, R56, R58, R60, R61, R65, R66, R67, R68, R69, R70, R71, R72, R73, R74, R75, R76, R77, R78, R79, R80, R81, R82, R87 \\ \hline
    Factor Measurement &   \\ \hline
    Percentage Based Measure & R1, R7, R12, R14, R21, R26, R29, R31, R33, R35, R36, R43, R46, R47, R49, R49, R53, R55, R56, R60, R61,R67, R71, R72, R73, R77, R78, R80 \\
    Gini & R15, R16 \\
    \hline
  \end{tabular}
\end{table*}

\subsubsection{Consensus Power Distribution} In the case of a Proof-of-Work based blockchain, the Consensus power is also known as the hash power of the miner (participating node). The centralization of hash power can pose a significant security threat to blockchain solutions such as Bitcoin and Ethereum. One key contributing factor to centralization is commercial mining pools. The income from mining operations depends on the probability of finding and propagating the solution of the puzzle before everyone else. The probability of successfully calculating the solution depends on the hash power of the computing device used for the calculation. Lower probability leads to a lack of stable income and may prompt users to mine as a group and share the profit. This group mining is also known as pooled mining \cite{lewenberg2015bitcoin}. Based on the analysis of the shortlisted literature, we report that the concept of pooled mining in itself is not considered a threat to the decentralization of the network; however, the literature is in agreement over the harms of a centrally run commercialized mining pool. In these centrally run mining pools, the pool manager decides which transactions to include in a block and subsequently distributes the workload among participants of the pool. This type of structure requires trusting the manager of the pool thus limiting the decentralization in the blockchain \cite{chesterman2018p2pool}. 
  
  \textbf{Measurement Technique }: Studies including \cite{SaiBuckleyLeGear2019,gencer2018decentralization,beikverdi2015trend,71}, have deployed an experimental setup to measure consensus centralization. Judmayer et al. (2017) \cite{71} refer to this approach as a ``block attribution scheme". In this experimental set-up, a participating node is connected to the blockchain that actively sniffs the network to extract mined blocks and coinbase addresses\footnote{The coin base address refers to the address of the node that gets the reward for successfully mining a block.}. The coin base address is then used to query public blockchain explorers to determine if it belongs to a known mining pool. Based on the results, a list of the mining pools and the proportion of the blocks mined by each respective public mining pool is constructed. Using this approach, we can calculate the proportion of total computational power that each mining pool controls.
  
This proportion can be represented as a percentage value as suggested by referred articles in Table \ref{tab:consensus} or by using the Gini values, based on the Lorenz Curve \cite{7,75}.

\textit{Gini Value Measurement:} These are economic measures of inequality \cite{gastwirth1971general,dorfman1979formula} for consensus power concentration \cite{7,75}.
  
  The \textit{Lorenz curve} is a graphical representation of the distribution of wealth. The curve illustrates the proportion of the income earned by any given percentage of the population. This curve has proven to be of significant importance in economic disparity measurement. To numerically describe this distribution, we can use the Gini Coefficient, which is based on the difference between the Lorenz curve and the line of equality \footnote{Line of equality refers to the equal distribution of hashing power among miners.}. We can calculate the Gini Coefficient as follows:
  
  \begin{equation}
    Gini = A/(A+B) 
\end{equation}

Where A is the area between the line of equality and Lorenz curve, and B is the area under the line of equality. The value of Gini can range between 0 to 1, where 0 represents complete equality, and 1 represents complete inequality.\\

\textbf{Implications of consensus power centralization}: The impact of centralization in consensus power has been widely studied in security literature \cite{chen2017security,zhang2019security,gervais2016security,karame2016security,SaiBuckleyLeGear2019,sapirshtein2016optimal}. A concentration of 26\% in proof of work-based blockchain can lead to successful selfish mining attacks. Whereas a consensus power concentration of over 51\% can result in a 51\% attack.

Smaller cryptocurrecies tend to be more prone to 51\% attack as evident by successful attacks on Aurum Coin, Bitcoin Gold, Ethereum Classic, Flo Blockchain, Monacoin, Verge, Vertcoin and ZenCash \cite{sayeed2019assessing}. These 51\% attacks have, on average, resulted in a loss of \$2.5 million per cryptocurrency \cite{sayeed2019assessing}. The significance of these attacks is evident by the agreement of all our interviewees on the centralization implications of a 51\% attack caused by consensus power concentration.

\subsection{Incentive Layer}

Bitcoin and similar decentralized cryptocurrencies are inherently dependent on the economics associated with rewards \cite{SaiBuckleyLeGear2019}. Sai et al. (2019) \cite{SaiBuckleyLeGear2019} reports that the exchange rate of Bitcoin is related to the overall consensus power of the network. If the exchange rate falls below a given threshold of profitability, the participants of the network may withdraw from active mining, which may result in a fall in overall hashing power of the network. A low value of hashing power of the network makes it easier for attackers to attain a higher consensus proportion; thus it may increase the threat of selfish mining and 51\% attack. This interplay between the monetary aspect of public cryptocurrencies and security makes it essential to inspect centralization on the economy driven incentive aspect of the network. A high concentration of wealth to a select few may be an aspect of centralization that can prove to be harmful to the network. Attacks such as the Whale Transaction Attack \cite{liao2017incentivizing} have exploited wealth concentration. In a whale transaction attack, the attacker attempts to induce disagreement\footnote{The disagreement is in the form of blockchain fork.} between the participants by providing a high transaction fee in an already published block. 

The issue of wealth concentration was raised by 60\% of our interview participants unprompted. $P_7$, for example, noted how they focused on wealth concentration:  \textit{``In general, I follow the money. If the trail of funds leads to one natural person or group of natural persons (regardless of number of addresses), then the process is relatively centralized along the spectrum of centralized-decentralized blockchain"}.

\begin{table}
    \centering
    \caption{Categories of centralization in Incentive Layer}
    \label{tab:incen}
  \begin{tabular}{|l|S|S|}
    \hline
    \multirow{2}{*}{Ref} &
      \multicolumn{2}{c|}{Wealth Concentration}   \\ \cline{2-3}
      & {Identification} & {Measurement}  \\
      \hline
    R16 & {\checkmark} & {$\times$}  \\
    R51 & {\checkmark} & {Gini} \\
    R52 & {\checkmark} & {Percentage Value}  \\
    R55 & {\checkmark} & {$\times$}  \\
    R62 & {\checkmark} & {Percentage Value}  \\
    R67 & {\checkmark} & {Gini}  \\
    \hline
  \end{tabular}
   
\end{table}

Table \ref{tab:incen} outlines the result of our review, identifying relevant articles and shortlisted techniques for measurement. In this subsection, we review the centralization based on Wealth Concentration in depth. This type of centralization may be of significance for a blockchain solution that employs a wealth-oriented consensus mechanism such as Proof-of-Stake \cite{kiayias2017ouroboros}.

 \subsubsection{Wealth Concentration} High accumulation of native cryptocurrency may give a unique advantage to an adversary. The high wealth concentration can also be used to increase the overall cost of transactions \cite{liao2017incentivizing}, as demonstrated in the iFish attack on the Ethereum network \cite{cryptoslate2018}. In the iFish attack, the attacker induced a large number of transactions with a high transaction fee in a short period. This influx of high transaction fees resulted in a considerable increase in the transaction fee. Another form of network abuse arising from high wealth concentration involves transaction fee manipulation by artificially increasing the overall fee required for a successful transaction. 
  
  Based on the results from our review, we point that this wealth concentration also has economic impacts on the network. As reported by Kondor et al. (2014) \cite{88}, already wealthy nodes in the bitcoin's transaction graph tend to increase their wealth at a higher speed than smaller nodes. They call this phenomena the \textit{``rich get richer"} scheme. 

  \textbf{Measurement Technique }: Wealth concentration measurement is at the center of disparity studies in economics \cite{gini1921measurement}. One of the most commonly used measures is the Gini Coefficient calculated from the Lorenz Curve. The wealth concentration is measured in the form of inequality based on the population and what proportion of population controls how much wealth. Translating this directly to the blockchain could mean calculating Gini over a cryptocurrency and all existing addresses on the blockchain. But we argue that this may not be the most efficient way as techniques such as Hierarchical Deterministic Wallets \cite{gutoski2015hierarchical} promote the generation of new addresses for every transaction. To overcome this limitation, Srinivasan et al. (2017) \cite{srinivasan2017} proposes establishing a lower bound value on the cryptocurrency contained in the address for inclusion in the measurement, i.e., a wallet with 0 cryptocurrencies may be excluded from the study, as it most likely resembles an inactive address. Another reported measurement technique is to use a percentage measure. However, a simple percentage measure fails to capture the distribution.

  \textbf{Implications of Wealth Concentration}: Wealth concentration is linked with a number of potential  attacks, such as the possibility of a 51\% attack in the case of a wealthy attacker during a fall in exchange rate \cite{SaiBuckleyLeGear2019}. Whale attack, as discussed above, is another example of a wealth oriented security threat to the network. However, both of these potential attacks are without any real-world incident reports. 
  
  One example of wealth concentration in a real-world attack is the transaction fee price manipulation caused by the iFish attack \cite{cryptoslate2018}. During the iFish attack, the attacker was able to artificially inflate the transaction fee of Ethereum by 35\%. Another example of a wealth oriented attack is the bZx hack, where a smart contract designed for lending Ether was exploited by sending high-value transactions and manipulating the platfrm\cite{zmudzinski2020}.   
  
  A public blockchain with high wealth concentration contradicts the foundational notion of a more even and open monetary system. This has a direct implication on the adoption of the technology.

\subsection{Operational Layer}
The uncertainty of reward imposes a constraint on participation for rational investors. This reasoning is primarily based on the cost of mining \cite{sai2019centralization}. A miner can earn rewards in the form of mining incentives and accumulated transaction fees from the mined block but to profitably mine on a Proof-of-Work blockchain, the difference between rewards earned and the expenses of the mining operation should be positive. This is the 'operations' we are referring to in this 'operational' layer. The expenses of mining operations include capital costs such as the acquisition of adequate hardware and other recurrent costs such as the cost of electricity. 

After conducting the systematic review, we report two types of centralization associated with operational aspects of the public blockchain. The first is the move from commercially available mining equipment to proprietary application-specific integrated circuit machines. This increased capital, operational cost has proven to be a significant barrier to entry for new miners in Bitcoin \cite{149}. We categorize this type of specialized hardware centralization as \textit{Specialized Equipment Concentration}.

Another factor that contributes to the cost of mining is the storage requirements for operating on the network. As all full nodes in the network are required to store and process all the transactions, the data stored increases \cite{dai2018low}. This imposes a significant barrier as traditional computing devices may not be able to participate in the network given high storage requirements. This may limit the participation in consensus to only the participants who can afford greater computational resources imposing a constraint on participation. A significant storage requirement may deter users with conventional computing devices from participating in the consensus altogether, resulting in a more centralized network converged on participants with high computational capabilities. This high storage requirement has been discussed as a centralization causing factor \cite{197,stopanddecrypt_2018,reddit}.  

In this layer of centralization, interviewee $I_{10}$ had an interesting perspective suggesting a restructuring of contract layer to widen our definition of the layer to include other operational concerns.

In this subsection, we report the centralization caused by the operational cost involved in participating in the consensus of the blockchain. We also manifest the result of our systematic literature review in Table \ref{tab:cenOperational}.

\begin{table}
    \centering
    \small
    \caption{Categories of centralization in Operational Layer}
    \label{tab:cenOperational}
  \begin{tabular}{|l|S|S|S|S|}
    \hline
    \multirow{2}{*}{Ref} &
      \multicolumn{2}{c}{Storage Constraint} &
      \multicolumn{2}{c|}{Specialized Equipment Concentration}  \\ \cline{2-5}
      & {Identification} & {Measurement} & {Identification} & {Measurement}  \\

      \hline
    R9 & {\checkmark} & {$\times$} & {$\times$} & {$\times$} \\
    R23 & {$\times$} & {$\times$} & {\checkmark}  & {$\times$} \\
    R24 & {\checkmark} & {Rate of Growth} & {$\times$} & {$\times$} \\
    R38 & {\checkmark} & {$\times$} & {$\times$} & {$\times$} \\
    R39 & {\checkmark} & {$\times$} & {$\times$} & {$\times$} \\
    R51 & {$\times$} & {$\times$} & {\checkmark} & {$\times$} \\
    R52 & {$\times$} & {$\times$} & {\checkmark}  & {$\times$} \\
    R53& {$\times$} & {$\times$} & {\checkmark} & {$\times$} \\
    R55& {$\times$} & {$\times$} & {\checkmark}  & {$\times$} \\
    R62& {$\times$} & {$\times$} & {\checkmark} & {$\times$} \\
    R63& {\checkmark} & {$\times$} & {$\times$} & {$\times$} \\
    R67& {$\times$} & {$\times$} & {\checkmark}  & {$\times$} \\
    R80& {\checkmark} & {$\times$} & {$\times$} & {$\times$} \\
    \hline
  \end{tabular}
   
\end{table}

\subsubsection{Size of the Blockchain} The traditional computing devices are often limited in-memory capabilities and can only hold a constrained amount of data. Attaining a higher storage capacity may prove to be costly if the growth rate of the storage requirement is significantly high \cite{197}. This growth in requirement may act as a deterring factor for non-organizational users as the requirement of the investment may be significant \cite{raman2017dynamic}, thus prompting centralization of mining effort. 
  
  The issue of storage requirement was articulated by 20\% of our interview participants. $I_{10}$ said: \textit{``Nothing really stops blockchains from becoming so large that we will run out of capacity. Personally, I have just experienced the first challenge because my Linux partition ran out of capacity; however, if I bought additional hard-disks, I will still be able to run a full node, but it is getting more expensive to run full nodes"}.

  \textbf{Measurement Technique }: To capture the storage-oriented centralization, Raman et al. (2017) \cite{raman2017dynamic} suggests using the growth rate as a metric. This growth rate is determined based on historical data about the total size of the blockchain. The growth rate can be calculated periodically, ideally after every difficulty recalibration \footnote{In Proof-of-Work based blockchains, the difficulty of finding the solution of the computational puzzle is updated after a predefined number of blocks to maintain a static block creation time.}. 
  
  $I_{10}$ stated their expectations for storage growth rate: \textit{``considering that Moore's Law applies to hard drives, it will be interesting to measure the growth rate in comparison with Moore’s law"}.
  
   \textbf{Implication of high storage requirements}: Every blockchain instance may have a different storage requirements, based on its implementation. For example, Bitcoin does not pose significant storage issues as the overall requirement is still low. In contrast, Ethereum has an important storage requirement where the growth rate may limit participation. A growing storage requirement for Ethereum may result in fewer people being able to participate in the network as the participating nodes on Ethereum are expected to store code of smart contracts. A low number of participating nodes increases the likelihood of a successful DDoS attack as it reduces the attack surface. 
  
  \subsubsection{Specialized Equipment Concentration} Proof-of-work based blockchains have seen a surge in the overall computational power of the network \cite{SaiBuckleyLeGear2019}. This surge has made it harder to get higher proportional control over consensus and, consequently, over the rewards associated with incentive. This higher computational requirement has induced an arms race in miners to acquire more efficient and specialized hardware \cite{22}. This type of specialized hardware is often not open source and gives the developers an advantage over others \cite{22}.

  60\% of our interview participants acknowledged specialized equipment concentration as an issue for a public blockchain. $I_7$ suggested that this concentration may undermine the whole proposition of public blockchains: \textit{``.. but blockchain doesn't live in a vacuum, so really it was/is the externalities (ASICs and other special hardware for example) that threw the biggest spanner in the experiment"}.

  \textbf{Measurement Technique }: Despite the significance of specialized equipment in Proof-of-work based mining operations, there is no existing metric to measure the centralization of hardware. Based on our literature review, we reason that this may be due to the non-public nature of this specialized hardware. As discussed earlier, most of these hardware implementations are not open source and often not available for public use.
  
  \textbf{Implication of Specialized Equipment Concentration}: As reported by several studies listed in Table \ref{tab:cenOperational}, the specialized equipment concentration may have given commercial entities an advantage over normal users. If this results in those commercial entries becoming focal, they may utilize the efficient computing equipment to attain higher consensus power and only release it to the public when it becomes less profitable to operate that computing equipment. This approach to hoarding efficient computing equipment is illustrated as the superhashing power dilemma by Bruschi et al. (2019) \cite{7}. As a result of our review, we suggest that further investigation is warranted into the measurement of specialized equipment and its impact on centralization. 
  
  Apart from the above reported DDoS attack due to the low number of nodes, the specialized equipment requirement severely contains the participation. This higher barrier of entry and lack of profitability with old hardware makes it impractical to contribute to the network without significant investment. This lack of involvement has been shown to increase the likelihood of a successful selfish mining and double-spending attack \cite{sai2019assessing}.

\subsection{Application Layer}
Users often rely on third-party applications to facilitate user interaction with the blockchain \cite{sai2019privacy}. These third-party applications include reference implementations, wallets, and exchanges \cite{gervais2014bitcoin}. As a result of our review, we report on centralization on these three application layer entities. We also suggest that a monopoly in the user end applications for a blockchain instance is a contributor to the centralization of the blockchain. This issue of centralization on third-party applications was also pointed out by $I_8$: \textit{``If you remember the catastrophe that centralized implementations such as Mt. Gox, Bitfinex have brought to the blockchain world, you can clearly see the desperate need for decentralization in user-facing applications"}. 

Results from our literature review are outlined in Table \ref{tab:appCent}.

\begin{table}
\tiny
    \centering
    \caption{Categories of centralization in Application Layer}
    \label{tab:appCent}
  \begin{tabular}{|l|S|S|S|S|S|S|}
    \hline
    \multirow{2}{*}{Ref} &
      \multicolumn{2}{c|}{Wallet Concentration} &
      \multicolumn{2}{c|}{Exchange Concentration}  &
      \multicolumn{2}{c|}{Reference Client} \\ \cline{2-7}
      & {Identification} & {Measurement} & {Identification} & {Measurement} & {Identification} & {Measurement} \\
     
      \hline
    R4 & {$\times$} & {$\times$} & {$\times$} & {$\times$} & {\checkmark} & {$\times$} \\
    R6 & {$\times$} & {$\times$} & {$\times$}& {$\times$} & {\checkmark} & {Satoshi Index} \\
    R8 & {$\times$} & {$\times$} & {$\times$} & {$\times$} & {$\times$} & {$\times$} \\
    R11 & {\checkmark} & {$\times$} & {\checkmark} & {$\times$} & {$\times$} & {$\times$}\\
    R13 & {\checkmark} & {$\times$} & {\checkmark} & {$\times$} & {$\times$} & {$\times$}\\
    R26 & {$\times$} & {$\times$} & {$\times$} & {$\times$}& {\checkmark} & {$\times$} \\
    R27 & {$\times$} & {$\times$} & {\checkmark} & {$\times$}& {$\times$}& {$\times$} \\
    R34 & {$\times$} & {$\times$} & {\checkmark} & {$\times$} & {$\times$} & {$\times$}\\
    R36 & {\checkmark} & {$\times$} & {$\times$} & {$\times$} & {\checkmark} & {$\times$}\\
    R37 & {$\times$} & {$\times$} & {\checkmark} & {Centrality} & {$\times$} & {$\times$}\\
    R40 & {\checkmark} & {$\times$} & {\checkmark} & {$\times$} & {$\times$} & {$\times$}\\
    R50 & {$\times$} & {$\times$} & {$\times$} & {$\times$} & {\checkmark} & {$\times$}\\
    R57 & {$\times$} & {$\times$} & {\checkmark} & {$\times$}& {$\times$} & {$\times$} \\
    R64 & {$\times$} & {$\times$} & {\checkmark} & {$\times$}& {$\times$} & {$\times$} \\
    R67 & {$\times$} & {$\times$} & {$\times$} & {$\times$} & {\checkmark} & {$\times$}\\
    R73 & {$\times$} & {$\times$} & {\checkmark} & {Centrality} & {$\times$} & {$\times$}\\
    R76 & {\checkmark} & {$\times$} & {$\times$} & {$\times$} & {$\times$} & {$\times$}\\
    R78 & {\checkmark} & {$\times$} & {\checkmark} & {$\times$} & {$\times$} & {$\times$}\\
    R83 & {$\times$} & {$\times$} & {$\times$} & {$\times$} & {\checkmark} & {$\times$}\\
    R84 & {\checkmark} & {$\times$} & {\checkmark} & {$\times$} & {$\times$} & {$\times$}\\
    R86 & {\checkmark} & {$\times$} & {\checkmark} & {$\times$} & {$\times$} & {$\times$}\\
    R88 & {\checkmark} & {$\times$} & {$\times$} & {$\times$} & {$\times$} & {$\times$}\\
    R89 & {$\times$} & {$\times$} & {\checkmark} & {$\times$} & {$\times$} & {$\times$}\\
    \hline
  \end{tabular}
   
\end{table}

This subsection is a manifestation of the identified centralization prone application layer entities. 

\subsubsection{Reference Client Development Concentration} As described in Section \ref{2.1background}, the data layer definition is implemented by a reference client, which acts as the gateway to the blockchain system. As any client that implements the protocol can become a part of the network, it is desirable from the decentralization point of view to have as many developers working on the reference implementation. Each client is expected to fulfill the protocol specification suggested by the core protocol. The development of the core protocol is decentralized by developing an open-source reference implementation. If a select few developers primarily drive the development of the core client, it contributes to centralization \cite{gervais2014bitcoin,azouvi2018egalitarian}. The decentralized protocol development factor captures this type of centralization. We note that this centralization is different from the improvement protocol centralization as the focus here on the development of a reference client and not improvements to the protocol. 
  
  Despite the reported adverse impact of this type of centralization on the blockchain, in Section \ref{discussion}, we present an argument in favor of some centralization in client development as the developer concentration may be a result of highly skilled developers making useful contributions. 
  
  \textbf{Measurement Technique }: \cite{gervais2014bitcoin} suggests examining the number of unique developers contributing to the open-source project with the number of commits on the main core client codebase. This approach is then extended by \cite{azouvi2018egalitarian}, where they propose using the Satoshi index, which represents the minimum percentage of all contributors required to reach 51\% of data contribution.  
  
   \textbf{Implication of reference client development concentration}: If only a select few developers work on the reference implementation, they may gain unfair influence over the network. This concentration of power in the hand of select few feeds into the governance issues discussed earlier. As discussed by \cite{gervais2014bitcoin,azouvi2018egalitarian}, this type of concentration is harmful to the decentralization of the network as a few developers may influence the implementation of change to the codebase. One of the major implications of influential actors in the public blockchain ecosystem is the defiance of open and equal monetary system assurance provided by the blockchain. As this open and equal system is one of the primary contributions of the public blockchain, the existence of influential entities severely limits systems capabilities to perform in an open and equal manner. 

\subsubsection{Exchange} Incentives for honest behavior are at the core of the decentralized, trustless transaction ledger. These incentives are often offered in the native cryptocurrency such as BTC and Ether. The real-world value of these cryptocurrencies has been debated \cite{SaiBuckleyLeGear2019} with the recommendation that they be determined by the exchange rate to traditional fiat currencies. The exchange of cryptocurrency to traditional fiat currency is aided by application layer entities known as exchanges. These exchanges act as the means of consensus formation around the exchange value. This process is also known as \textit{Price Discovery}. Due to the vital importance of the exchanges, the exchange applications must not be monopolized. 
  
  \textbf{Measurement Technique }: To measure the state of centralization in exchanges, Marvin et al. (2017) propose measuring the centrality of exchanges by examining the flow of cryptocurrencies between addresses on the blockchain \cite{marvin2017decentralised}. Addresses with high centrality in transactions may point to exchanges. This is observed by graphing the transaction flow and identifying nodes with a high degree of centrality. This was followed by the calculation of a Gini Coefficient that reports on the trend of centralization due to exchanges. 
  
 Other studies, such as Hileman et al. (2017) \cite{123}, have employed a percentage-based value measure, where they measure the proportion of all bitcoin transactions processed by exchanges.
  
   \textbf{Implication of centralized exchanges}: A large number of successful attacks on Bitcoin and Ethereum have focused on exploiting vulnerabilities in exchanges \cite{chia2018rethinking}. These centralized systems act as a single point of failure in case they also serve as a central repository of keys. A prominent example of this is the closure of Mt. Gox due to numerous security flaws leading to loss of Bitcoins owned by its users \cite{abrams2014erosion}. 
   
  Attacks on centralized exchanges not only impact the users of the exchange but the broader cryptocurrency community as it can instill doubts over the security of the ecosystem. These security attacks contribute to the barring trust and adoption by the wider community.
   
\subsubsection{Wallet Concentration} Wallet applications are another form of centralized service on the application layer, as these applications are often developed and maintained by centralized organizations \cite{sai2019privacy}.   

  \textbf{Measurement Technique}: Based on the review of the relevant literature, we report that there are no suggestions regarding the measurement of wallet concentration. We reason that this may be due to the nature of how wallets operate in a closed commercial environment. However, as most of these wallets use an exchange service to transmit funds such as Coinbase \cite{123}, it may be reasoned that exchange centralization may provide a rough proxy for wallet based centralization as well.  
  
   \textbf{Implication of centralized wallet}: Applications such as wallets have been identified as a single point of failure and are considered a security threat \cite{sai2019privacy}. A high concentration of wealth in centrally managed wallets may give the host an advantage feeding into the issue of wealth concentration. This concentration may also result in a dependence on centralized organization, consequently reducing the decentralization. 
 
 Similar to exchanges, a centralized wallet poses a potential barrier of entry in the ecosystem. Due to the technical ability required to host their wallets, most end users tend to prefer hosted wallets, which provides attackers with a small attack surface. This can aid attackers in conducting more targeted yet profitable attacks on the centralized wallet hosting service.

\section{State of centralization in Bitcoin and Ethereum}

The following subsection provides an overview of empirical evidence specific to the two most prominently used blockchain-based cryptocurrencies: Bitcoin and Ethereum. We present the view of the literature on the centralization of these two cryptocurrencies. To structure this investigation, we use the initial taxonomy. The results from this investigation are manifested in Table \ref{tab:centralizationInBitcoinAndEthereum}.

\begin{table*}
\centering
  \caption{State of centralization in Bitcoin and Ethereum}
  \label{tab:centralizationInBitcoinAndEthereum}
  \begin{tabular}{|p{5cm}|p{4cm}|p{4cm}|}
    \hline
    Centralization Factor & Bitcoin & Ethereum \\
    \hline
   Wallet Concentration &  No Measurement &  No Measurement\\ \hline
    Exchange Concentration & 7 exchanges served more than 95\% of all trades & Data not available  \\ \hline
    Reference Client Concentration & Single developer authored 30\% of all files &  Single developer authored 55\% of all files \\ \hline
                       
   Storage Growth Rate & 0.1-0.5 GB per day  & Data not available  \\ \hline
   Specialized Equipment Concentration & No Measurement  & No Measurement  \\ \hline                    
   Wealth Concentration & Gini = 0.65 & Gini = 0.76 \\ \hline
                    
    Consensus Power Distribution & Top 4 mining pools with 53\% consensus power & Top 3 mining pools with 61\% consensus power \\ \hline
    
    Node Discovery Protocol Control & No Measurement  & No Measurement \\ \hline
     Geographic Distribution &  Network latency 26.7\% less than Ethereum & 26.7\% higher than Bitcoin \\ \hline
    Bandwidth Concentration & 1.9 to 2.7 times greater than Ethereum  & 1.9 to 2.7 times less than Bitcoin \\ \hline
    Routing Centralization  & 30\% network with 10 ASes  & 28\% network with 1 AS \\ \hline
    Owner Control  & $C_{OwnerControl}$ = 0.033 & $C_{OwnerControl}$ = 0.11 \\ \hline
    Improvement Protocol & Mean = 11.41, Median = 2.0, IQR = 6.5, IQMean = 2.95 & Mean = 9.16, Median = 2.0, IQR = 5.0, IQMean = 2.56    
 \\ \hline

  \end{tabular}
\end{table*}

\subsection{Governance Layer} 
\textbf{Owner Control}: Satoshi Nakamoto is largely credited for the authorship and development of bitcoin \cite{nakamoto2008bitcoin}. This as-yet unknown individual or organization is said to have performed active mining in the early days of Bitcoin, accumulating a considerable amount of BTC \cite{bohr2014uses}. As Bitcoin implements provisions for anonymity, the amount of BTC held by Satoshi is not publicly known. Based on the data obtained from the Bitcoin blocks mined in 2009 \cite{bitcoinss}, \cite{bitslog_2013} performed clustering of similar wallets to identify large entities gathering BTCs. They also report that the largest gain of around 700,000 BTC belonged to a single entity performing mining in 2009. This gives us a value of $C_{OwnerControl}$= 0.033 for Bitcoin. 

This value is significantly less than that of Ethereum, where the value of $C_{OwnerControl}$ is 0.11. Bai et al. (2020) \cite{bai2020evolution} argues that Ethereum is very unfair since ``the rich are already very rich". This high wealth disparity may allow select participants to conduct attacks based on economic manipulation of the Ethereum ecosystem, such as the Whale transaction attack and transaction fee manipulation. A reason for the high value has been presented in Section \ref{4taxonomy}.

\textbf{Improvement Protocol}: According to \cite{bitcoin_2019}, all changes must be approved by all the developers of the core client. However, Gervais et al. (2014) \cite{gervais2014bitcoin} reported on one violation of this rule. In this violation, a subset of developers unilaterally decided to lower the minimum transaction fee to 0.0001 BTC. This illustration strengthens \cite{gervais2014bitcoin} questioning of the transparency, in the process of handling improvement proposals. 

Azouvi et al. (2018) \cite{azouvi2018egalitarian} measures the centralization by calculating the centrality metrics for both Bitcoin and Ethereum reported in Table \ref{tab:centralizationInBitcoinAndEthereum}. They report that the collected commits and comments data set contained many outliers pointing out that the top 25 \% of developers contribute significantly more than others. They also point out that in their data set for Ethereum, a vast majority of EIP contributions are from a single user, Vitalik Buterin, the founder of Ethereum. They report a similar trend in Bitcoin, where only a handful of people are contributing to the improvement protocol allowing these select groups of people to dictate the changes that are implemented in the protocol. In the past, the block size debate surrounding the scalability has often been cited as a prime example of this type of governance control. 

\subsection{Network Layer}
\textbf{Node Discovery Protocol Control}: As reported in Section \ref{4taxonomy}, our literature review suggest that no study has focused on the measurement of centralization of seed DNS nodes. We postulate that further investigation is required to measure this type of centralization adequately. 

\textbf{Geographic Distribution}: Gencer et al. (2018) \cite{gencer2018decentralization} conducted an extensive review of both Bitcoin and Etherum to measure centralization in the network layer. They reported that the Bitcoin network is more geographically centralized than Ethereum. The average peer-to-peer network latency of Ethereum is 26.7\% higher than Bitcoin, suggesting that Ethereum nodes are located at a greater geographic distance. They reason that this is due to the data center focused approach to mining for Bitcoin, whereas Ethereum can be mined by using consumer hardware. This association between geographical distribution and operational centralization neatly illustrates the interdependency between different aspects of centralization, even those based in different layers.

\textbf{Bandwidth Concentration}: Gencer et al. (2018) \cite{gencer2018decentralization} states that nodes in Bitcoin tend to have about 1.9 to 2.7 times more network bandwidth than Ethereum nodes. They also report that based on the bandwidth, it can be assumed that Bitcoin nodes are located in data center clusters, whereas Ethereum exhibits a more spread out distribution of bandwidth.

\textbf{Routing Centralization}: Feld et al. (2014) \cite{feld2014analyzing} reports that 30\% of the bitcoin network was only made up of 10 ASes, which presents a level of security threat. This work was expanded by Apostolaki et al. (2017) \cite{apostolaki2017hijacking}, where they report that 13 ASes covered about 30\% of the network but only consisted of 36 IP prefixes. These 36 IP prefixes cover about 50\% of mining power. However, the only investigation that has reported on AS-Level centralization in Ethereum, Gencer et al. (2018) \cite{gencer2018decentralization} reports that 28\% of Ethereum nodes belonged to a single AS.

\subsection{Consensus Layer}
Centralization of consensus power of bitcoin has been studied thoroughly in the literature \cite{beikverdi2015trend,gervais2016security,gervais2014bitcoin,SaiBuckleyLeGear2019,karame2016bitcoin}. Beikverdi et al. (2015) \cite{beikverdi2015trend} uses a percentage based centralization value to derive a new metric called Centralization Factor. They report that at the beginning of 2011, 30\% of all hashing power was controlled by eight mining pools. This concentration sees a significant increase in 2014 when, according to Gervais et al. (2014) \cite{gervais2014bitcoin}, the top mining pool alone controls close to 40 \% of all hashing power of the network. 

Gencer et al. (2018) \cite{gencer2018decentralization} expands these analyses by also examining Ethereum's network. During the observation period, Gencer et al. (2018) \cite{gencer2018decentralization} reports that Bitcoin had a less centralized consensus mechanism than Ethereum. On average, the top four mining pools in Bitcoin controlled 53\% of the hashing power, whereas in Ethereum the top three mining pools controlled 61\% hashing power. 

\subsection{Incentive Layer}
According to Malik et al. (2016) \cite{malik2016history}, as of 2016, 11,000 unique Bitcoin addresses, out of a total of 12 million, contained 75.2\% of all Bitcoin in circulation. This disparity shows a significant concentration of wealth to a select few. Chohan (2019) \cite{chohan2019cryptocurrencies} also supports the claim of significant inequality in the Bitcoin network. The author claims that the level of inequality reflects that of traditional economies and voids the proposed purpose of Bitcoin: decentralization. Gupta et al. (2017) \cite{gupta2017gini} conducted an in-depth investigation of the inequality of Bitcoin. They report that Bitcoin had a Gini value of 0.995 in the year 2013. This result is then refined by Srinivasan et al. (2017) \cite{srinivasan2017}, where they set a lower bound on the Bitcoin account to account for Hierarchical Deterministic wallets as described in Section \ref{3methodology}. They report that in 2018, Bitcoin had a Gini value of 0.65, where they set the minimum threshold to 185 BTC per account. This Gini value suggests that wealth in bitcoin is highly centralized when compared to real economies where, according to the World Bank \cite{worldbank}, the highest reported Gini value is 0.63.

According to Srinivasan et al. (2017) \cite{srinivasan2017}, Ethereum demonstrates a similar trend of significant centralization with a Gini value of 0.76 with a minimum threshold of 2477 ETH per account. This suggested trend is in line with the report by \cite{insightsvol}, where they claim Ethereum to be more centralized in terms of wealth distribution.

\subsection{Operational Layer}
In Pustivsek et al. (2019) \cite{pustivsek2019approaching}, the authors report that the Bitcoin full node requires 204 GB storage space. This storage requirement is slightly lower than the 385 GB required by Ethereum for a full node \cite{afanasev2018design}. Pustivsek et al. (2019) \cite{pustivsek2019approaching} also reports that the storage growth rate is about 0.1-0.5 GB per day. Our review was unable to identify any longitudinal studies that observe the growth in storage requirements over a long time.

As reported in Section \ref{4taxonomy}, numerous studies identify specialized equipment concentration as a cause of centralization. Despite the significant attention to this issue, our review suggests that there are no proposed measurement techniques.

\subsection{Application Layer}
\textbf{Reference Client Concentration}: According to Azouvi et al. (2018) \cite{azouvi2018egalitarian}, a single author wrote about 30\% of all files in the bitcoin reference implimentation \footnote{Azouvi et al. (2018) \cite{azouvi2018egalitarian} propose using the Satoshi Index to measure centralization in client development. However, the specific values of Satoshi Index for Bitcoin and Ethereum are not available.}. This is significantly higher in Ethereum, where an individual author wrote 55\% of all files. They also analyze the comments on the GitHub pages of Bitcoin and Ethereum reference clients. They report that only eight people contributed to half of all comments representing 0.3\% of all commenters. This concentration in comments is also observable in Ethereum, where 0.6\% commenters contributed to 50\% of comments. 

\textbf{Exchange Concentration}: Intermediary services such as Exchanges that also act as central key stores for Bitcoin have been suggested as a centralization causing factor by Bohme et al. (2015) \cite{bohme2015bitcoin}. A prominent example of the harm caused by exchange concentration is the collapse of Mt. Gox in 2014 \cite{abrams2014erosion}. In 2014, Mt. Gox was the leading exchange for Bitcoin, and its closure resulted in a total loss of \$450 Million. Bohme et al. (2015) \cite{bohme2015bitcoin} reports that the concentration of exchanges was still high in 2015 when the seven largest exchanges served more than 95\% of all bitcoin trades. 

An empirical analysis conducted by Bohme et al. (2015) \cite{bohme2015bitcoin} reported that out of 40 Bitcoin exchanges examined, 18 had closed, wiping out customers' account balance as they stored the private keys of customers. They argue that these exchanges operate as the de facto centralized authorities in the Bitcoin network. 

As for Ethereum, we report that there are no studies that explicitly report on the behavior of exchanges for Ethereum. However, as suggested by Kim et al. (2018) \cite{kim2018risk}, most of the Bitcoin exchanges also exchange multiple other cryptocurrencies, including Ether.

As discussed earlier, based on our systematic review, we conclude that there is no suggestion regarding a measurement technique to capture wallet based centralization. 

So, in terms of Bitcoin, the main centralization threats are at the Network, Consensus, and Application layers. Specifically, the centralization aspects of the Network layer: geographic distribution, bandwidth, and routing are vulnerabilities for bitcoin in that they allow the specific threats of geopolitical manipulation of the network, high resource requirement for participation, and possibility of network attacks. These threats for bitcoin are augmented by the high concentration of consensus power to centralized mining pools and application layer operations such as exchanges and wallets.

Ethereum also shares the issues of centralization on the application layer as they lead to reliance on centralized entities such as exchanges and wallets for participation in the network. Other significant centralization threats for Ethereum include the Governance, Consensus, and Incentive layers. Especially the centralization aspects of the Governance and Incentive layers may induce vulnerabilities for Ethereum in that they allow unilateral decision making on the governance layer and high wealth concentration on the incentive layer.   

\section{Discussion}\label{discussion}


In this first in-depth investigation of the centralization of public blockchain solutions, we conducted a systematic review of existing literature to produce an initial taxonomy of centralization. We then refined this initial taxonomy through expert interviews. We provide an overview of centralization in different aspects of the blockchain. We examine different means of measuring centralization, also pointing out the absence of measurement techniques in these research studies. This initial taxonomy provides a framework for a more systematic discussion around the centralization of major blockchain systems. The following section discusses the findings of our survey.

\subsection{Non Binary Nature of Centralization}

We observe that decentralization in the public Blockchain literature is a loosely-defined term that can take many shapes and forms. We also observe that most of the non-decentralization-specific articles reviewed treat decentralization as a binary construct. That is: a blockchain instance is either centralized or decentralized. However, based on our taxonomy, we define centralization of public Blockchains as \textit{the process by which one or more architectural dimensions (aspects) of the Blockchain are restrictive to the majority of participants by direct or indirect economic, social, or technical constraints} and so argue that centralization is not suited to binary classification.

This latter observation aligns with expert interviews, where 60\% of participants preferred a spectrum of values for centralization rather than the conventional binary notion. However, the interviewees also acknowledged that the complexity of a more granular definition might dilute the meaning to non-experts in the blockchain domain. For example, $I_5$ said:  \textit{``I am an engineer, so I prefer precision and a multidimensional model, but I know when you are presenting to business people, a single score might be what they are looking for"}. 

This survey presents a novel, initial taxonomy to address this dilution concern and allows for structured discussion on centralization. The following text discusses the key findings of the taxonomy.

Consensus power concentration was the most recognized form of blockchain centralization by both the literature and experts interviewed. We reason that this wide recognition is due to the dependence of significant security threats such as the Double Spending \cite{karame2012double} and Selfish mining \cite{sapirshtein2016optimal} attacks on the consensus power concentration.  The practical implication of this centralization is the heavy the impact of mining pools when operating a profitable mining operation. The dominance of mining pools is observable in both Ethereum and Bitcoin. In Bitcoin the top 4 mining pools control over 53\% of the hashing power, whereas in Ethereum the top 3 mining pools control over 61\% of the hashing power (See Table \ref{tab:centralizationInBitcoinAndEthereum}).

A high concentration of consensus power can induce an arm's race to attain the most efficient hardware \cite{SaiBuckleyLeGear2019}. Our survey reports that this race often results in specialized proprietary hardware. The practical implication of this type of hardware concentration is an indirect limitation to participation as only efficient, and often proprietary hardware can result in a profitable operation. To remedy this situation, studies such as Cho et al. Hyungmin (2018) \cite{42}, have proposed using a consensus algorithm that is memory heavy, for which specialized hardware design is inefficient.

Surprisingly on a similar operational constraint, the Storage growth rate was less widely recognized to contribute to centralization. However, $I_{10}$ raised an interesting issue on the ever-increasing append-only nature of Blockchain that may result in consistent growth in storage requirements. As reported in Table \ref{tab:centralizationInBitcoinAndEthereum}, the current growth rate for Bitcoin is around 0.1 to 0.5 GB per day. The practical implication of this increased storage requirement is the inability of conventional computing devices to serve as nodes in the blockchain \cite{197}. Guo et al. \cite{197} propose a storage optimization scheme based on the redundant residual number system that can reduce the storage requirement. We suggest that a further investigation into storage optimization in public Blockchain is warranted.

Another unexpected finding of our survey was that 50\% of the interviewees accepted node discovery protocol control as a threat to decentralization, despite only one research article reporting on the issue.  We reason that this may be due to the practical implications of setting up a new node such as the potential delay in network connection for new nodes due to high traffic through DNS nodes. This type of delay is often not accounted for in network simulation tools such as NS3, employed by studies such as \cite{gervais2016security, SaiBuckleyLeGear2019}.

Contrary to the previous example, routing and bandwidth centralization in the network was not widely recognized by the interviewees. One potential explanation could be the experimental nature of the measurement associated with the routing and bandwidth centralization. Despite these being recognized as issues, both the bandwidth and routing do not cause operational issues to most participants at present. 

Another network-oriented centralization concern widely recognized by both the literature and interviewees is the geographic distribution of the nodes. Our findings suggest that the Ethereum network is more geographically spread out than Bitcoin. We reason that this is due to the possibility of using conventional hardware such as GPUs to participate in Ethereum. Despite the recognition, our literature review did not identify potential strategies to address this centralization. We suggest that strategies to limit geographic concentration should be investigated. 

The lack of mitigation techniques is also persistent in the application layer aspects. The wallet and exchange centralization have been reported on by the literature and also recognized as centralization issues by expert interviews. As reasoned earlier, the centralized store of cryptocurrencies may give an advantage to the exchange or wallet operator. This advantage is often in the form of wealth concentration and can be observed in the centralization of Bitcoin exchange platforms, where only seven exchanges were reported to serve more than 95\% of all trades. 

Interviewees and literature also agree on the implication of wealth concentration on the decentralization. Surprisingly, despite the apparent issue of a \textit{``Rich getting Richer"} effect in Proof-of-Stake cryptocurrencies \cite{fanti2019compounding}, most of the reported literature focused on the wealth concentration in Proof-of-Work. We suggest that the issue of wealth concentration be investigated in the context of Proof-of-Stake cryptocurrencies. 

Another factor that may result in a \textit{``Rich getting Richer"} effect is the distribution of wealth at the very start of the Blockchain captured by owner control in our taxonomy. The issue of owner control is also associated with how the Blockchain is governed. Governance centralization in Blockchain is widely recognized by both the literature and interviewees. Interestingly, Wang et al. (2017) \cite{wang2017internal} argue for some centralization in the governance to facilitate quick response to security threats. We expand on this line of reasoning in the following subsection.

\subsection{Aspect based Measurement of Implications of Centralization}

As pointed out earlier, not all aspects of our taxonomy are an equal contributor to the overall centralization of the blockchain. This was also substantiated by six interviewees agreeing that a combined value of centralization for the overall blockchain would not be meaningful. For example, storage constraint oriented centralization may be an issue in Ethereum due to the requirement to store smart contracts. In contrast, this may not be a significant issue for Bitcoin as only transactions drive the storage requirements. We expand on this category-based significance reasoning that not all centralization is necessarily equally bad for the network: 

The governance layer based centralization argument presented by Gervais et al. (2014) \cite{gervais2014bitcoin} assumes that concentrating decision making power to a select few is bad for blockchain. However, we question this argument, as true decentralization is an impossibility in real world scenarios \cite{szabo1970,kwon2019impossibility}. The concentration in decision making had also proven to be useful in instances of network attacks when a prompt response was mandated \cite{wang2017internal}. Delegation of controlling power during the cases of security bugs or attacks may have proven to be detrimental to the network. Despite the lack of decentralization in governance, it may be to the overall benefit of the network. We present this as a potential future research avenue to explore the most suitable governance structure for decentralized systems. 

We also argue that the results obtained by Azouvi et al. (2018) \cite{azouvi2018egalitarian} regarding the centralization in source code development for core client implementation may not necessarily be bad. It may just be the case that only a handful of developers have an in-depth understanding of the source code to make useful contributions to the system. This reasoning of limited expertise feeds into the argument against the decentralization of the improvement protocol. As pointed out by Azouvi et al. (2018) \cite{azouvi2018egalitarian}, the vast majority of the Ethereum Improvement Protocol recommendations originated from a single developer, Vitalik Buterin. We reason that this may be due to the quality of suggestions proposed by Vitalik. 

These arguments in favor of some centralization are an example of the complex nature of decentralization in distributed systems. We propose that the significance of each aspect of centralization be determined based on the empirical evidence specific to each blockchain instance.

\section{Conclusion}
In this paper, we conduct a systematic literature review to provide a summary of the research done on the centralization aspect of blockchain. We structure our findings in a novel initial taxonomy of centralization. This taxonomy is then refined and validated through expert interviews.

\subsection{Contribution}

Decentralized blockchain solutions provide a means of monetary asset transfer without a trusted third party; this is attained through the delegation of the validation power to all participants of the system rather than the administrator. This delegation of control is often referred to as the original contribution of blockchain systems \cite{bonneau2015sok}. Based on previous studies, \cite{SaiBuckleyLeGear2019,gencer2018decentralization,cong2019decentralized,gervais2014bitcoin}, we reason that the preconceived notion that blockchains are inherently decentralized may not hold in the present situation and that raises the potential of severe issues for blockchain instances. Due to the lack of an objective measure of centralization, it becomes impractical to discuss improvement in terms of centralization.

Centralization is a challenging variable to research, in part because of the multiple definitions and measures of centralization applicable in blockchain and, to date, the implicit nature of several of those aspects and the lack of an encompassing framework. We report on these myriads of definitions, conceptualizations, and dimensions used to describe this concept by segmenting them based on a generic architecture proposed by Zhang et al. (2019) \cite{zhang2019security}. Our study contributes to the existing body of knowledge by systematically surveying and synthesizing the blockchain literature, reporting on the adverse impact of centralization such as security threats, as well as identifying research gaps such as the lack of Ethereum specific research on centralization.

With this systematic review, we provide the reader with an overview of various forms of centralization in Blockchain resulting in an initial taxonomy. This taxonomy also contains numerous existing measurement techniques used to measure centralization. It may help researchers evaluate the centralization of a blockchain instance, but will also allow researchers add more aspects of centralization as they become known, providing them with a vocabulary of centralization that will allow them address the issues that arise.

We have also reported on the platform-specific findings for the two most prominently used blockchain-based cryptocurrencies: Bitcoin and Ethereum. We report that both Bitcoin and Ethereum have similar centralization issues with regards to reference client implementation, decentralized protocol development, and exchanges. However, in terms of wealth concentration, Ethereum is more centralized than Bitcoin, primarily due to high owner control. This trend continues with consensus power concentration, where Ethereum is reported to be more centralized than Bitcoin. Ethereum nodes, however, are geographically more spread out than Bitcoin, resulting in a low geographic concentration when compared to Bitcoin.  

We also discuss that centralization on all aspects is not necessarily adverse for the blockchain by expanding the argument in favor of some centralization by Wang et al. (2017) \cite{wang2017internal}. We suggest that the unpropitious impact of centralization be measured on each aspect based on empirical evidence. This aspect-specific investigation may assist the move from the binary notion of decentralization to a multidimensional scale encompassing adequate measurement and control where necessary. 


\subsection{Threats to validity \label{8.1}}

As decentralization is fundamental to a public blockchain, the term is frequently used in the title and abstract of articles relating to public blockchains. To not omit any relevant articles, we kept the search queries generic by including any article that includes the term \textit{``Blockchain"} and \textit{``Decentralization"} along with suggested alternate words in Section \ref{3methodology}. We acknowledge that despite the broad terms used, we may have missed relevant articles not present, or with different phrasing, on these leading search repositories. These missed articles may include \textit{``grey literature"}, which is of significant importance in the blockchain research domain \cite{casino2019systematic}. To overcome this limitation, we included Google Scholar in our search process. However, as reported earlier, the Google Scholar search was limited to the top 1,000 entries, even though the relevant articles dropped off significantly after the top four hundred returned articles.

The literature review may also be limited due to the strict inclusion and exclusion criteria for the title and abstract filtering. We reason that these strict criteria are warranted due to a large number of articles retrieved by the search queries (3,574 non-duplicate entries). To overcome this limitation, we employed a two-step filtration by reviewing both the title and abstract. We also performed cross-validation of the filtration process by the independent review of the articles by two authors. This cross-validation process resulted in Cohen's Kappa value of 0.84, which is considered an almost perfect agreement. We repeated a similar cross-validation process for the full-text filtration. 

The review process aimed to extract factors from all shortlisted articles despite their core focus. As the study of centralization in public blockchain is still in the early stage, we included articles where the core focus was not centralization. This inclusion may have limited the quality of shortlisted articles, as observed by the exclusion of 148 articles after full-text filtration. To overcome this limitation, we performed a quality review of all 212 shortlisted articles and shortlisted a final set of 89 articles.

To further evaluate the literature-review findings, we interviewed ten experts. The recruitment process was based on the prominence of authors in the bibliographic map generated by Ramona et al. (2019) \cite{ramona2019bitcoin}. As with any other qualitative research method, interviews have several limitations, as pointed out by Opdenakker et al. (2006) \cite{opdenakker2006advantages}. In addressing them, we adher to the validity dimensions put forth by Maxwell (1992) \cite{maxwell1992understanding} for qualitative studies. The first validity threat is the descriptive validity of the data obtained through interviews. To limit this, we transcribed the audio-captured interview in verbatim form. However, in the interviews that relied on contemporaneous notes, it is possible that the interviewer may have missed some observations. The second threat to validity is the interpretive validity of the interviews. To address this, we used open-ended questions and restricted the questions strictly to the research questions presented in Section \ref{3.2methodology}. We also coded the interviews based on the terms used by the interviewees rather than an interpretation. The transcripts and notes were individually checked by researchers from the author list. The interviewees were also given back the interview transcripts and notes for validation.

\subsection{Future Work}

Having provided a comprehensive overview of centralization in public blockchain, a case study focused on individual cryptocurrencies, and blockchain implementations would complement our study. This case study could include an in-depth centralization review of, for example, Bitcoin, Ethereum, and Libra \cite{pilkington2019libra}. 

The taxonomy developed by our study can also be expanded to provide an objective measure of centralization for blockchain instances, as a whole, to facilitate comparison. This objective measure may prove to be useful for the evaluation of centralization from a novice user, or governance perceptive. Four of our ten interviewees stated that they would prefer a single score to measure centralization objectively, and thought it would assist end-users and nonspecialist researchers.

We also hope to develop different flavors of this initial taxonomy that are specific to implementation details. For instance, the presented taxonomy is generic and does not consider consensus specific issues such as Stake bleeding \cite{gavzi2018stake}. It also omits the consideration of source code dependencies in Smart Contracts. In future, we intend to statistically examine the source code of smart contracts to observe if a handful of libraries dominate the smart contracts in Ethereum.

The work presented here only examines the already identified factors that may lead to centralization and does not analyze the existence of other novel forms of centralization. As a part of future work, we will consider a thorough review of one of the reference blockchain implementations to identify factors that may also contribute to centralization directly or indirectly. 

We also aim to review existing literature to identify potential solutions to the centralization avenues suggested by our review. These solutions may facilitate integrating centralization considerations during the development of public blockchains.



\section*{Acknowledgements}
This  work  was  supported,  in part,  by  Science  Foundation  Ireland  grant 13/RC/2094 and co-funded under the European Regional Development Fund through the Southern  Eastern Regional Operational Programme to Lero - the Irish Software Research Centre (www.lero.ie).

\
\bibliographystyle{unsrt}  
\bibliography{references}  

\begin{thebibliography}{100}

\bibitem{beck2017blockchain}
Roman Beck, Michel Avital, Matti Rossi, and Jason~Bennett Thatcher.
\newblock Blockchain technology in business and information systems research,
  2017.

\bibitem{yli2016current}
Jesse Yli-Huumo, Deokyoon Ko, Sujin Choi, Sooyong Park, and Kari Smolander.
\newblock Where is current research on blockchain technology?---a systematic
  review.
\newblock {\em PLoS ONE}, 11:e0163477, 2016.

\bibitem{mattila2016blockchain}
Juri Mattila.
\newblock The blockchain phenomenon--the disruptive potential of distributed
  consensus architectures.
\newblock Technical report, ETLA working papers, 2016.

\bibitem{androulaki2018hyperledger}
Elli Androulaki, Artem Barger, Vita Bortnikov, Christian Cachin, Konstantinos
  Christidis, Angelo De~Caro, David Enyeart, Christopher Ferris, Gennady
  Laventman, Yacov Manevich, et~al.
\newblock Hyperledger fabric: a distributed operating system for permissioned
  blockchains.
\newblock In {\em Proceedings of the Thirteenth EuroSys Conference}, page~30.
  ACM, 2018.

\bibitem{wust2018you}
Karl W{\"u}st and Arthur Gervais.
\newblock Do you need a blockchain?
\newblock In {\em 2018 Crypto Valley Conference on Blockchain Technology
  (CVCBT)}, pages 45--54. IEEE, 2018.

\bibitem{zheng2017overview}
Zibin Zheng, Shaoan Xie, Hongning Dai, Xiangping Chen, and Huaimin Wang.
\newblock An overview of blockchain technology: Architecture, consensus, and
  future trends.
\newblock In {\em 2017 IEEE International Congress on Big Data (BigData
  Congress)}, pages 557--564. IEEE, 2017.

\bibitem{beck2018governance}
Roman Beck, Christoph M{\"u}ller-Bloch, and John~Leslie King.
\newblock Governance in the blockchain economy: A framework and research
  agenda.
\newblock {\em Journal of the Association for Information Systems},
  19(10):1020--1034, 2018.

\bibitem{walport2016distributed}
Mark Walport.
\newblock Distributed ledger technology: beyond blockchain. uk government
  office for science.
\newblock Technical report, Tech. Rep, 2016.

\bibitem{davidson2016economics}
Sinclair Davidson, Primavera De~Filippi, and Jason Potts.
\newblock Economics of blockchain.
\newblock {\em Available at SSRN 2744751}, 2016.

\bibitem{he2017survey}
Pu~He, Ge~Yu, YF~Zhang, and YB~Bao.
\newblock Survey on blockchain technology and its application prospect.
\newblock {\em Computer Science}, 44(4):1--7, 2017.

\bibitem{cong2019decentralized}
Lin~William Cong, Zhiguo He, and Jiasun Li.
\newblock Decentralized mining in centralized pools.
\newblock Technical report, National Bureau of Economic Research, 2019.

\bibitem{gervais2014bitcoin}
Arthur Gervais, Ghassan~O Karame, Vedran Capkun, and Srdjan Capkun.
\newblock Is bitcoin a decentralized currency?
\newblock {\em IEEE security \& privacy}, 12(3):54--60, 2014.

\bibitem{SaiBuckleyLeGear2019}
Ashish~Rajendra Sai, Jim Buckley, and Andrew Le~Gear.
\newblock Assessing the security implication of bitcoin exchange rates.
\newblock {\em Computers and Security}, Jun 2019.

\bibitem{gencer2018decentralization}
Adem~Efe Gencer, Soumya Basu, Ittay Eyal, Robbert Van~Renesse, and Emin~G{\"u}n
  Sirer.
\newblock Decentralization in bitcoin and ethereum networks.
\newblock {\em arXiv preprint arXiv:1801.03998}, 2018.

\bibitem{beikverdi2015trend}
Alireza Beikverdi and JooSeok Song.
\newblock Trend of centralization in bitcoin's distributed network.
\newblock In {\em 2015 IEEE/ACIS 16th International Conference on Software
  Engineering, Artificial Intelligence, Networking and Parallel/Distributed
  Computing (SNPD)}, pages 1--6. IEEE, 2015.

\bibitem{azouvi2018egalitarian}
Sarah Azouvi, Mary Maller, and Sarah Meiklejohn.
\newblock Egalitarian society or benevolent dictatorship: The state of
  cryptocurrency governance.
\newblock In {\em International Conference on Financial Cryptography and Data
  Security}, pages 127--143. Springer, 2018.

\bibitem{kwon2019impossibility}
Yujin Kwon, Jian Liu, Minjeong Kim, Dawn Song, and Yongdae Kim.
\newblock Impossibility of full decentralization in permissionless blockchains.
\newblock {\em arXiv preprint arXiv:1905.05158}, 2019.

\bibitem{zhang2019security}
Rui Zhang, Rui Xue, and Ling Liu.
\newblock Security and privacy on blockchain.
\newblock {\em ACM Computing Surveys (CSUR)}, 52(3):1--34, 2019.

\bibitem{judmayer2017blocks}
Aljosha Judmayer, Nicholas Stifter, Katharina Krombholz, and Edgar Weippl.
\newblock Blocks and chains: Introduction to bitcoin, cryptocurrencies, and
  their consensus mechanisms.
\newblock {\em Synthesis Lectures on Information Security, Privacy, \& Trust},
  9(1):1--123, 2017.

\bibitem{baliga2017understanding}
Arati Baliga.
\newblock Understanding blockchain consensus models.
\newblock {\em Persistent}, 2017(4):1--14, 2017.

\bibitem{104}
Wenbo Wang, Dinh~Thai Hoang, Zehui Xiong, Dusit Niyato, Ping Wang, Peizhao Hu,
  and Yonggang Wen.
\newblock A survey on consensus mechanisms and mining management in blockchain
  networks.
\newblock {\em arXiv preprint arXiv:1805.02707}, pages 1--33, 2018.

\bibitem{43}
Naif Alzahrani and Nirupama Bulusu.
\newblock Towards true decentralization: A blockchain consensus protocol based
  on game theory and randomness.
\newblock In {\em International Conference on Decision and Game Theory for
  Security}, pages 465--485. Springer, 2018.

\bibitem{bonneau2015sok}
Joseph Bonneau, Andrew Miller, Jeremy Clark, Arvind Narayanan, Joshua~A Kroll,
  and Edward~W Felten.
\newblock Sok: Research perspectives and challenges for bitcoin and
  cryptocurrencies.
\newblock In {\em 2015 IEEE Symposium on Security and Privacy}, pages 104--121.
  IEEE, 2015.

\bibitem{guegan:halshs-01524440}
Dominique Guegan.
\newblock {Public Blockchain versus Private blockhain}, April 2017.
\newblock Documents de travail du Centre d'Economie de la Sorbonne 2017.20 -
  ISSN : 1955-611X.

\bibitem{meijer2018governance}
David Meijer and Jolien Ubacht.
\newblock The governance of blockchain systems from an institutional
  perspective, a matter of trust or control?
\newblock In {\em Proceedings of the 19th Annual International Conference on
  Digital Government Research: Governance in the Data Age}, pages 1--9, 2018.

\bibitem{peck2017blockchain}
Morgen~E Peck.
\newblock Blockchain world-do you need a blockchain? this chart will tell you
  if the technology can solve your problem.
\newblock {\em IEEE Spectrum}, 54(10):38--60, 2017.

\bibitem{sai2019privacy}
Ashish~Rajendra Sai, Jim Buckley, and Andrew Le~Gear.
\newblock Privacy and security analysis of cryptocurrency mobile applications.
\newblock In {\em 2019 Fifth Conference on Mobile and Secure Services
  (MobiSecServ)}, pages 1--6. IEEE, 2019.

\bibitem{karame2012double}
Ghassan~O Karame, Elli Androulaki, and Srdjan Capkun.
\newblock Double-spending fast payments in bitcoin.
\newblock In {\em Proceedings of the 2012 ACM conference on Computer and
  communications security}, pages 906--917, 2012.

\bibitem{halpin2017introduction}
Harry Halpin and Marta Piekarska.
\newblock Introduction to security and privacy on the blockchain.
\newblock In {\em 2017 IEEE European Symposium on Security and Privacy
  Workshops (EuroS\&PW)}, pages 1--3. IEEE, 2017.

\bibitem{karame2016bitcoin}
Ghassan~O Karame and Elli Androulaki.
\newblock {\em Bitcoin and blockchain security}.
\newblock Artech House, 2016.

\bibitem{karame2016security}
Ghassan Karame.
\newblock On the security and scalability of bitcoin's blockchain.
\newblock In {\em Proceedings of the 2016 ACM SIGSAC conference on computer and
  communications security}, pages 1861--1862. ACM, 2016.

\bibitem{sapirshtein2016optimal}
Ayelet Sapirshtein, Yonatan Sompolinsky, and Aviv Zohar.
\newblock Optimal selfish mining strategies in bitcoin.
\newblock In {\em International Conference on Financial Cryptography and Data
  Security}, pages 515--532. Springer, 2016.

\bibitem{iansiti2017truth}
Marco Iansiti and Karim~R Lakhani.
\newblock The truth about blockchain.
\newblock {\em Harvard Business Review}, 95(1):118--127, 2017.

\bibitem{conti2018survey}
Mauro Conti, E~Sandeep Kumar, Chhagan Lal, and Sushmita Ruj.
\newblock A survey on security and privacy issues of bitcoin.
\newblock {\em IEEE Communications Surveys \& Tutorials}, 20(4):3416--3452,
  2018.

\bibitem{chohan2019cryptocurrencies}
Usman~W Chohan.
\newblock Cryptocurrencies and inequality.
\newblock {\em Notes on the 21st Century (CBRI)}, 2019.

\bibitem{wang2017internal}
Sha Wang, Jean-Philippe~JP Vergne, and Ying-Ying Hsieh.
\newblock The internal and external governance of blockchain-based
  organizations: Evidence from cryptocurrencies.
\newblock In {\em Bitcoin and beyond}, pages 48--68. Routledge, 2017.

\bibitem{caffyn2015bitcoin}
Grace Caffyn.
\newblock What is the bitcoin block size debate and why does it matter.
\newblock {\em URL: http://www. coindesk.
  com/what-is-the-bitcoin-block-size-debate-and-why-does-it-matter/(visited on
  27/11/2015)}, 2015.

\bibitem{wirdum2016rejecting}
Aaron~van Wirdum.
\newblock Rejecting today’s hard fork, the ethereum classic project continues
  on the original chain: Here’s why.
\newblock {\em Bitcoin Magazine, July}, 20, 2016.

\bibitem{nakamoto2008bitcoin}
Satoshi Nakamoto et~al.
\newblock Bitcoin: A peer-to-peer electronic cash system.
\newblock 2008.

\bibitem{kitchenham2004procedures}
Barbara Kitchenham.
\newblock Procedures for performing systematic reviews.
\newblock 2004.

\bibitem{great2016distributed}
Great Britain. Government~Office for Science.
\newblock {\em Distributed Ledger Technology: Beyond Block Chain}.
\newblock Government Office for Science, 2016.

\bibitem{o2014bitcoin}
Karl~J O'Dwyer and David Malone.
\newblock Bitcoin mining and its energy footprint.
\newblock 2014.

\bibitem{35}
Cong~T Nguyen, Dinh~Thai Hoang, Diep~N Nguyen, Dusit Niyato, Huynh~Tuong
  Nguyen, and Eryk Dutkiewicz.
\newblock Proof-of-stake consensus mechanisms for future blockchain networks:
  fundamentals, applications and opportunities.
\newblock {\em IEEE Access}, 7:85727--85745, 2019.

\bibitem{mingxiao2017review}
Du~Mingxiao, Ma~Xiaofeng, Zhang Zhe, Wang Xiangwei, and Chen Qijun.
\newblock A review on consensus algorithm of blockchain.
\newblock In {\em 2017 IEEE International Conference on Systems, Man, and
  Cybernetics (SMC)}, pages 2567--2572. IEEE, 2017.

\bibitem{guo2016blockchain}
Ye~Guo and Chen Liang.
\newblock Blockchain application and outlook in the banking industry.
\newblock {\em Financial Innovation}, 2(1):24, 2016.

\bibitem{panarello2018blockchain}
Alfonso Panarello, Nachiket Tapas, Giovanni Merlino, Francesco Longo, and
  Antonio Puliafito.
\newblock Blockchain and iot integration: A systematic survey.
\newblock {\em Sensors}, 18(8):2575, 2018.

\bibitem{zhu2019applications}
Qingyi Zhu, Seng~W Loke, Rolando Trujillo-Rasua, Frank Jiang, and Yong Xiang.
\newblock Applications of distributed ledger technologies to the internet of
  things: A survey.
\newblock {\em ACM Computing Surveys (CSUR)}, 52(6):1--34, 2019.

\bibitem{dwivedi2019decentralized}
Ashutosh~Dhar Dwivedi, Gautam Srivastava, Shalini Dhar, and Rajani Singh.
\newblock A decentralized privacy-preserving healthcare blockchain for iot.
\newblock {\em Sensors}, 19(2):326, 2019.

\bibitem{xie2019survey}
Junfeng Xie, Helen Tang, Tao Huang, F~Richard Yu, Renchao Xie, Jiang Liu, and
  Yunjie Liu.
\newblock A survey of blockchain technology applied to smart cities: Research
  issues and challenges.
\newblock {\em IEEE Communications Surveys \& Tutorials}, 21(3):2794--2830,
  2019.

\bibitem{li2017survey}
Xiaoqi Li, Peng Jiang, Ting Chen, Xiapu Luo, and Qiaoyan Wen.
\newblock A survey on the security of blockchain systems.
\newblock {\em Future Generation Computer Systems}, 2017.

\bibitem{bradbury2013problem}
Danny Bradbury.
\newblock The problem with bitcoin.
\newblock {\em Computer Fraud \& Security}, 2013(11):5--8, 2013.

\bibitem{de2019fragility}
Manlio De~Domenico and Andrea Baronchelli.
\newblock The fragility of decentralised trustless socio-technical systems.
\newblock {\em EPJ Data Science}, 8(1):2, 2019.

\bibitem{garay2015bitcoin}
Juan Garay, Aggelos Kiayias, and Nikos Leonardos.
\newblock The bitcoin backbone protocol: Analysis and applications.
\newblock In {\em Annual International Conference on the Theory and
  Applications of Cryptographic Techniques}, pages 281--310. Springer, 2015.

\bibitem{gervais2016security}
Arthur Gervais, Ghassan~O Karame, Karl W{\"u}st, Vasileios Glykantzis, Hubert
  Ritzdorf, and Srdjan Capkun.
\newblock On the security and performance of proof of work blockchains.
\newblock In {\em Proceedings of the 2016 ACM SIGSAC conference on computer and
  communications security}, pages 3--16, 2016.

\bibitem{briscoe2000understanding}
Neil Briscoe.
\newblock Understanding the osi 7-layer model.
\newblock {\em PC Network Advisor}, 120(2), 2000.

\bibitem{antonopoulos2018mastering}
Andreas~M Antonopoulos and Gavin Wood.
\newblock {\em Mastering ethereum: building smart contracts and dapps}.
\newblock O'reilly Media, 2018.

\bibitem{antonopoulos2017mastering}
Andreas~M Antonopoulos.
\newblock {\em Mastering bitcoin: Programming the open blockchain}.
\newblock " O'Reilly Media, Inc.", 2017.

\bibitem{buterin2013ethereum}
Vitalik Buterin et~al.
\newblock Ethereum white paper.
\newblock {\em GitHub repository}, pages 22--23, 2013.

\bibitem{wood2014ethereum}
Gavin Wood et~al.
\newblock Ethereum: A secure decentralised generalised transaction ledger.
\newblock {\em Ethereum project yellow paper}, 151(2014):1--32, 2014.

\bibitem{lee2019using}
Wei-Meng Lee.
\newblock Using the web3. js apis.
\newblock In {\em Beginning Ethereum Smart Contracts Programming}, pages
  169--198. Springer, 2019.

\bibitem{chu2018broker}
Dennis Chu.
\newblock Broker-dealers for virtual currency: Regulating cryptocurrency
  wallets and exchanges.
\newblock {\em Columbia Law Review}, 118(8):2323--2360, 2018.

\bibitem{petersen2008systematic}
Kai Petersen, Robert Feldt, Shahid Mujtaba, and Michael Mattsson.
\newblock Systematic mapping studies in software engineering.
\newblock In {\em 12th International Conference on Evaluation and Assessment in
  Software Engineering (EASE) 12}, pages 1--10, 2008.

\bibitem{galster2013variability}
Matthias Galster, Danny Weyns, Dan Tofan, Bartosz Michalik, and Paris Avgeriou.
\newblock Variability in software systems—a systematic literature review.
\newblock {\em IEEE Transactions on Software Engineering}, 40(3):282--306,
  2013.

\bibitem{razzaq2018state}
Abdul Razzaq, Asanka Wasala, Chris Exton, and Jim Buckley.
\newblock The state of empirical evaluation in static feature location.
\newblock {\em ACM Transactions on Software Engineering and Methodology
  (TOSEM)}, 28(1):1--58, 2018.

\bibitem{fleiss1973equivalence}
Joseph~L Fleiss and Jacob Cohen.
\newblock The equivalence of weighted kappa and the intraclass correlation
  coefficient as measures of reliability.
\newblock {\em Educational and psychological measurement}, 33(3):613--619,
  1973.

\bibitem{sim2005kappa}
Julius Sim and Chris~C Wright.
\newblock The kappa statistic in reliability studies: use, interpretation, and
  sample size requirements.
\newblock {\em Physical therapy}, 85(3):257--268, 2005.

\bibitem{landis1977measurement}
J~Richard Landis and Gary~G Koch.
\newblock The measurement of observer agreement for categorical data.
\newblock {\em biometrics}, pages 159--174, 1977.

\bibitem{ramona2019bitcoin}
Or{\u{a}}ștean Ramona, M{\u{a}}rginean~Silvia Cristina, Sava Raluca, et~al.
\newblock Bitcoin in the scientific literature--a bibliometric study.
\newblock {\em Studies in Business and Economics}, 14(3):160--174, 2019.

\bibitem{122}
Nouriel Roubini.
\newblock The big blockchain lie.
\newblock {\em Project Syndicate. Blog post. Oct}, 15, 2018.

\bibitem{87}
Marcella Atzori.
\newblock Blockchain technology and decentralized governance: Is the state
  still necessary?
\newblock {\em Available at SSRN 2709713}, 2015.

\bibitem{wolfson2015bitcoin}
Shael~N Wolfson.
\newblock Bitcoin: the early market.
\newblock {\em Journal of Business \& Economics Research (JBER)},
  13(4):201--214, 2015.

\bibitem{ethereumtransactioninformation}
Etherscan.
\newblock Ethereum transaction information, 2019.

\bibitem{totalETH}
Etherscan.
\newblock Total ether supply and market capitalization, 2019.

\bibitem{anceaume2016safety}
Emmanuelle Anceaume, Thibaut Lajoie-Mazenc, Romaric Ludinard, and Bruno
  Sericola.
\newblock Safety analysis of bitcoin improvement proposals.
\newblock In {\em 2016 IEEE 15th International Symposium on Network Computing
  and Applications (NCA)}, pages 318--325. IEEE, 2016.

\bibitem{de2016invisible}
Primavera De~Filippi and Benjamin Loveluck.
\newblock The invisible politics of bitcoin: governance crisis of a
  decentralized infrastructure.
\newblock {\em Internet Policy Review}, 5(4), 2016.

\bibitem{bitcoin_2019}
Bitcoin.
\newblock bitcoin/bips, Jun 2019.

\bibitem{kim2019ethics}
Tae~Wan Kim and Ariel Zetlin-Jones.
\newblock The ethics of blockchain networks] the ethics of contentious hard
  forks in blockchain networks with fixed features.
\newblock {\em Frontiers in Blockchain}, 2:9, 2019.

\bibitem{58}
Seoung~Kyun Kim, Zane Ma, Siddharth Murali, Joshua Mason, Andrew Miller, and
  Michael Bailey.
\newblock Measuring ethereum network peers.
\newblock In {\em Proceedings of the Internet Measurement Conference 2018},
  pages 91--104, 2018.

\bibitem{roubini_2018}
Nouriel Roubini.
\newblock Blockchain isn't about democracy and decentralisation – it's about
  greed | nouriel roubini, Oct 2018.

\bibitem{36}
Till Neudecker and Hannes Hartenstein.
\newblock Network layer aspects of permissionless blockchains.
\newblock {\em IEEE Communications Surveys \& Tutorials}, 21(1):838--857, 2018.

\bibitem{apostolaki2017hijacking}
Maria Apostolaki, Aviv Zohar, and Laurent Vanbever.
\newblock Hijacking bitcoin: Routing attacks on cryptocurrencies.
\newblock In {\em 2017 IEEE Symposium on Security and Privacy (SP)}, pages
  375--392. IEEE, 2017.

\bibitem{miller2015discovering}
Andrew Miller, James Litton, Andrew Pachulski, Neal Gupta, Dave Levin, Neil
  Spring, and Bobby Bhattacharjee.
\newblock Discovering bitcoin’s public topology and influential nodes.
\newblock {\em et al}, 2015.

\bibitem{tapsell2018evaluation}
James Tapsell, Raja~Naeem Akram, and Konstantinos Markantonakis.
\newblock An evaluation of the security of the bitcoin peer-to-peer network.
\newblock In {\em 2018 IEEE International Conference on Internet of Things
  (iThings) and IEEE Green Computing and Communications (GreenCom) and IEEE
  Cyber, Physical and Social Computing (CPSCom) and IEEE Smart Data
  (SmartData)}, pages 1057--1062. IEEE, 2018.

\bibitem{jin2017blockndn}
Tong Jin, Xiang Zhang, Yirui Liu, and Kai Lei.
\newblock Blockndn: A bitcoin blockchain decentralized system over named data
  networking.
\newblock In {\em 2017 Ninth International Conference on Ubiquitous and Future
  Networks (ICUFN)}, pages 75--80. IEEE, 2017.

\bibitem{dietrich2000analyzing}
Sven Dietrich, Neil Long, and David Dittrich.
\newblock Analyzing distributed denial of service tools: The shaft case.
\newblock In {\em LISA}, pages 329--339, 2000.

\bibitem{heilman2015eclipse}
Ethan Heilman, Alison Kendler, Aviv Zohar, and Sharon Goldberg.
\newblock Eclipse attacks on bitcoin’s peer-to-peer network.
\newblock In {\em 24th $\{$USENIX$\}$ Security Symposium ($\{$USENIX$\}$
  Security 15)}, pages 129--144, 2015.

\bibitem{54}
Irni~Eliana Khairuddin and Corina Sas.
\newblock An exploration of bitcoin mining practices: Miners' trust challenges
  and motivations.
\newblock In {\em Proceedings of the 2019 CHI Conference on Human Factors in
  Computing Systems}, pages 1--13, 2019.

\bibitem{saroiu2001measurement}
Stefan Saroiu, P~Krishna Gummadi, and Steven~D Gribble.
\newblock Measurement study of peer-to-peer file sharing systems.
\newblock In {\em Multimedia Computing and Networking 2002}, volume 4673, pages
  156--170. International Society for Optics and Photonics, 2001.

\bibitem{19}
Zibin Zheng, Shaoan Xie, Hong-Ning Dai, Xiangping Chen, and Huaimin Wang.
\newblock Blockchain challenges and opportunities: A survey.
\newblock {\em International Journal of Web and Grid Services}, 14(4):352--375,
  2018.

\bibitem{feld2014analyzing}
Sebastian Feld, Mirco Sch{\"o}nfeld, and Martin Werner.
\newblock Analyzing the deployment of bitcoin's p2p network under an as-level
  perspective.
\newblock {\em Procedia Computer Science}, 32:1121--1126, 2014.

\bibitem{lewenberg2015bitcoin}
Yoad Lewenberg, Yoram Bachrach, Yonatan Sompolinsky, Aviv Zohar, and Jeffrey~S
  Rosenschein.
\newblock Bitcoin mining pools: A cooperative game theoretic analysis.
\newblock In {\em Proceedings of the 2015 International Conference on
  Autonomous Agents and Multiagent Systems}, pages 919--927. International
  Foundation for Autonomous Agents and Multiagent Systems, 2015.

\bibitem{chesterman2018p2pool}
Xavier Chesterman.
\newblock {\em THE P2POOL MINING POOL}.
\newblock PhD thesis, Ghent University, 2018.

\bibitem{71}
Aljosha Judmayer, Alexei Zamyatin, Nicholas Stifter, Artemios~G Voyiatzis, and
  Edgar Weippl.
\newblock Merged mining: Curse or cure?
\newblock In {\em Data Privacy Management, Cryptocurrencies and Blockchain
  Technology}, pages 316--333. Springer, 2017.

\bibitem{7}
Francesco Bruschi, Vincenzo Rana, Lorenzo Gentile, and Donatella Sciuto.
\newblock Mine with it or sell it: the superhashing power dilemma.
\newblock {\em ACM SIGMETRICS Performance Evaluation Review}, 46(3):127--130,
  2019.

\bibitem{75}
Fabio Caccioli, Giacomo Livan, and Tomaso Aste.
\newblock Scalability and egalitarianism in peer-to-peer networks.
\newblock In {\em Banking beyond banks and money}, pages 197--212. Springer,
  2016.

\bibitem{gastwirth1971general}
Joseph~L Gastwirth.
\newblock A general definition of the lorenz curve.
\newblock {\em Econometrica: Journal of the Econometric Society}, pages
  1037--1039, 1971.

\bibitem{dorfman1979formula}
Robert Dorfman.
\newblock A formula for the gini coefficient.
\newblock {\em The review of economics and statistics}, pages 146--149, 1979.

\bibitem{chen2017security}
Lin Chen, Lei Xu, Nolan Shah, Zhimin Gao, Yang Lu, and Weidong Shi.
\newblock On security analysis of proof-of-elapsed-time (poet).
\newblock In {\em International Symposium on Stabilization, Safety, and
  Security of Distributed Systems}, pages 282--297. Springer, 2017.

\bibitem{sayeed2019assessing}
Sarwar Sayeed and Hector Marco-Gisbert.
\newblock Assessing blockchain consensus and security mechanisms against the
  51\% attack.
\newblock {\em Applied Sciences}, 9(9):1788, 2019.

\bibitem{liao2017incentivizing}
Kevin Liao and Jonathan Katz.
\newblock Incentivizing blockchain forks via whale transactions.
\newblock In {\em International Conference on Financial Cryptography and Data
  Security}, pages 264--279. Springer, 2017.

\bibitem{kiayias2017ouroboros}
Aggelos Kiayias, Alexander Russell, Bernardo David, and Roman Oliynykov.
\newblock Ouroboros: A provably secure proof-of-stake blockchain protocol.
\newblock In {\em Annual International Cryptology Conference}, pages 357--388.
  Springer, 2017.

\bibitem{cryptoslate2018}
Cryptoslate.
\newblock Ethereum network under assault: Gas price manipulation may indicate
  covert eos attack [interview], Sep 2018.

\bibitem{88}
D{\'a}niel Kondor, M{\'a}rton P{\'o}sfai, Istv{\'a}n Csabai, and G{\'a}bor
  Vattay.
\newblock Do the rich get richer? an empirical analysis of the bitcoin
  transaction network.
\newblock {\em PloS one}, 9(2), 2014.

\bibitem{gini1921measurement}
Corrado Gini.
\newblock Measurement of inequality of incomes.
\newblock {\em The Economic Journal}, 31(121):124--126, 1921.

\bibitem{gutoski2015hierarchical}
Gus Gutoski and Douglas Stebila.
\newblock Hierarchical deterministic bitcoin wallets that tolerate key leakage.
\newblock In {\em International Conference on Financial Cryptography and Data
  Security}, pages 497--504. Springer, 2015.

\bibitem{srinivasan2017}
Balaji~S. Srinivasan.
\newblock Quantifying decentralization, Oct 2017.

\bibitem{zmudzinski2020}
Adrian Zmudzinski.
\newblock Decentralized lending protocol bzx hacked twice in a matter of days,
  Feb 2020.

\bibitem{sai2019centralization}
Ashish~Rajendra Sai, Andrew Le~Gear, and Jim Buckley.
\newblock Centralization threat metric.
\newblock 2019.

\bibitem{149}
Maria Borge, Eleftherios Kokoris-Kogias, Philipp Jovanovic, Linus Gasser,
  Nicolas Gailly, and Bryan Ford.
\newblock Proof-of-personhood: Redemocratizing permissionless cryptocurrencies.
\newblock In {\em 2017 IEEE European Symposium on Security and Privacy
  Workshops (EuroS\&PW)}, pages 23--26. IEEE, 2017.

\bibitem{dai2018low}
Mingjun Dai, Shengli Zhang, Hui Wang, and Shi Jin.
\newblock A low storage room requirement framework for distributed ledger in
  blockchain.
\newblock {\em IEEE Access}, 6:22970--22975, 2018.

\bibitem{197}
Zhaohui Guo, Zhen Gao, Haojuan Mei, Ming Zhao, and Jinsheng Yang.
\newblock Design and optimization for storage mechanism of the public
  blockchain based on redundant residual number system.
\newblock {\em IEEE Access}, 7:98546--98554, 2019.

\bibitem{stopanddecrypt_2018}
StopAndDecrypt.
\newblock The ethereum-blockchain size has exceeded 1tb, and yes, it's an
  issue, May 2018.

\bibitem{reddit}
Reddit.
\newblock r/monero - blockchain size issue in future?, 2019.

\bibitem{raman2017dynamic}
Ravi~Kiran Raman and Lav~R Varshney.
\newblock Dynamic distributed storage for scaling blockchains.
\newblock {\em arXiv preprint arXiv:1711.07617}, 2017.

\bibitem{22}
Ariel Ekblaw, Chelsea Barabas, Jonathan Harvey-Buschel, and Andrew Lippman.
\newblock Bitcoin and the myth of decentralization: Socio-technical proposals
  for restoring network integrity.
\newblock In {\em 2016 IEEE 1st International Workshops on Foundations and
  Applications of Self* Systems (FAS* W)}, pages 18--23. IEEE, 2016.

\bibitem{sai2019assessing}
Ashish~Rajendra Sai, Jim Buckley, and Andrew Le~Gear.
\newblock Assessing the security implication of bitcoin exchange rates.
\newblock {\em Computers \& Security}, 2019.

\bibitem{marvin2017decentralised}
Ian Marvin.
\newblock {\em Decentralised? A Study of Concentration in the Bitcoin Network.}
\newblock PhD thesis, University of Cape Town, 2017.

\bibitem{123}
Garrick Hileman and Michel Rauchs.
\newblock Global cryptocurrency benchmarking study.
\newblock {\em Cambridge Centre for Alternative Finance}, 33, 2017.

\bibitem{chia2018rethinking}
Vincent Chia, Pieter Hartel, Qingze Hum, Sebastian Ma, Georgios Piliouras,
  Daniel Reijsbergen, Mark van Staalduinen, and Pawel Szalachowski.
\newblock Rethinking blockchain security: Position paper.
\newblock {\em arXiv preprint arXiv:1806.04358}, 2018.

\bibitem{abrams2014erosion}
Rachel Abrams, Matthew Goldstein, and Hiroko Tabuchi.
\newblock Erosion of faith was death knell for mt. gox.
\newblock {\em New York Times}, 2014.

\bibitem{bohr2014uses}
Jeremiah Bohr and Masooda Bashir.
\newblock Who uses bitcoin? an exploration of the bitcoin community.
\newblock In {\em 2014 Twelfth Annual International Conference on Privacy,
  Security and Trust}, pages 94--101. IEEE, 2014.

\bibitem{bitcoinss}
Blockchain luxembourg s.a.
\newblock Blocks mined, 2019.

\bibitem{bitslog_2013}
Sergio.
\newblock The well deserved fortune of satoshi nakamoto, bitcoin creator,
  visionary and genius, Apr 2013.

\bibitem{bai2020evolution}
Qianlan Bai, Chao Zhang, Yuedong Xu, Xiaowei Chen, and Xin Wang.
\newblock Evolution of ethereum: A temporal graph perspective.
\newblock {\em arXiv preprint arXiv:2001.05251}, 2020.

\bibitem{malik2016history}
Vladimir Malik.
\newblock The history and the future of bitcoin.
\newblock {\em Praha: Bankovn{\'\i} institut vysok{\'a} {\v{s}}kola Praha},
  2016.

\bibitem{gupta2017gini}
Manas Gupta and Parth Gupta.
\newblock Gini coefficient based wealth distribution in the bitcoin network: A
  case study.
\newblock In {\em International Conference on Computing, Analytics and
  Networks}, pages 192--202. Springer, 2017.

\bibitem{worldbank}
World Bank.
\newblock Gini index (world bank estimate).

\bibitem{insightsvol}
Huobi Blockchain Big Data~Weekly Insights.
\newblock Vol. 8.

\bibitem{pustivsek2019approaching}
Matev{\v{z}} Pusti{\v{s}}ek, Anton Umek, and Andrej Kos.
\newblock Approaching the communication constraints of ethereum-based
  decentralized applications.
\newblock {\em Sensors}, 19(11):2647, 2019.

\bibitem{afanasev2018design}
Maxim~Ya Afanasev, Anastasiya~A Krylova, Sergey~A Shorokhov, Yuri~V Fedosov,
  and Anastasiia~S Sidorenko.
\newblock A design of cyber-physical production system prototype based on an
  ethereum private network.
\newblock In {\em 2018 22nd Conference of Open Innovations Association
  (FRUCT)}, pages 3--11. IEEE, 2018.

\bibitem{bohme2015bitcoin}
Rainer B{\"o}hme, Nicolas Christin, Benjamin Edelman, and Tyler Moore.
\newblock Bitcoin: Economics, technology, and governance.
\newblock {\em Journal of economic Perspectives}, 29(2):213--38, 2015.

\bibitem{kim2018risk}
Chang~Yeon Kim and Kyungho Lee.
\newblock Risk management to cryptocurrency exchange and investors guidelines
  to prevent potential threats.
\newblock In {\em 2018 International Conference on Platform Technology and
  Service (PlatCon)}, pages 1--6. IEEE, 2018.

\bibitem{42}
Hyungmin Cho.
\newblock Asic-resistance of multi-hash proof-of-work mechanisms for blockchain
  consensus protocols.
\newblock {\em IEEE Access}, 6:66210--66222, 2018.

\bibitem{fanti2019compounding}
Giulia Fanti, Leonid Kogan, Sewoong Oh, Kathleen Ruan, Pramod Viswanath, and
  Gerui Wang.
\newblock Compounding of wealth in proof-of-stake cryptocurrencies.
\newblock In {\em International Conference on Financial Cryptography and Data
  Security}, pages 42--61. Springer, 2019.

\bibitem{szabo1970}
Nick Szabo.
\newblock The dawn of trustworthy computing, Jan 1970.

\bibitem{casino2019systematic}
Fran Casino, Thomas~K Dasaklis, and Constantinos Patsakis.
\newblock A systematic literature review of blockchain-based applications:
  current status, classification and open issues.
\newblock {\em Telematics and Informatics}, 36:55--81, 2019.

\bibitem{opdenakker2006advantages}
Raymond Opdenakker.
\newblock Advantages and disadvantages of four interview techniques in
  qualitative research.
\newblock In {\em Forum qualitative sozialforschung/forum: Qualitative social
  research}, volume~7, 2006.

\bibitem{maxwell1992understanding}
Joseph Maxwell.
\newblock Understanding and validity in qualitative research.
\newblock {\em Harvard educational review}, 62(3):279--301, 1992.

\bibitem{pilkington2019libra}
Marc Pilkington.
\newblock The libra project: A transnational monetary dystopia--analysis of the
  disruption generated by the facebook-led stable coin.
\newblock {\em Available at SSRN 3434079}, 2019.

\bibitem{gavzi2018stake}
Peter Ga{\v{z}}i, Aggelos Kiayias, and Alexander Russell.
\newblock Stake-bleeding attacks on proof-of-stake blockchains.
\newblock In {\em 2018 Crypto Valley Conference on Blockchain Technology
  (CVCBT)}, pages 85--92. IEEE, 2018.

\end{thebibliography}

\end{document}